\documentclass[10pt,twocolumn,letterpaper]{article}

\usepackage[pagenumbers]{iccv} 

\definecolor{iccvblue}{rgb}{0.21,0.49,0.74}
\usepackage[pagebackref,breaklinks,colorlinks,allcolors=iccvblue]{hyperref}

\usepackage{amssymb}
\makeatletter

\newcommand{\Rmnum}[1]{\expandafter\@slowromancap\romannumeral #1@}
\makeatother
\usepackage{graphicx}
\usepackage{multirow}
\usepackage{amsthm,amsmath,amssymb}
\usepackage{mathrsfs}
\usepackage{makecell}
\usepackage{bm}

\def\confName{ICCV}
\def\confYear{2025}

\title{StableCodec: Taming One-Step Diffusion for Extreme Image Compression}

\author{Tianyu Zhang, Xin Luo, Li Li, Dong Liu \\
University of Science and Technology of China, Hefei, China \\
{\tt\small \{zhangtianyu, xinluo\}@mail.ustc.edu.cn, \{lil1, dongeliu\}@ustc.edu.cn} 
}

\begin{document}

\twocolumn[{
	\renewcommand\twocolumn[1][]{#1}%
	\maketitle
	\begin{center}
		\centering
		\captionsetup{type=figure}
		\includegraphics[width=\linewidth]{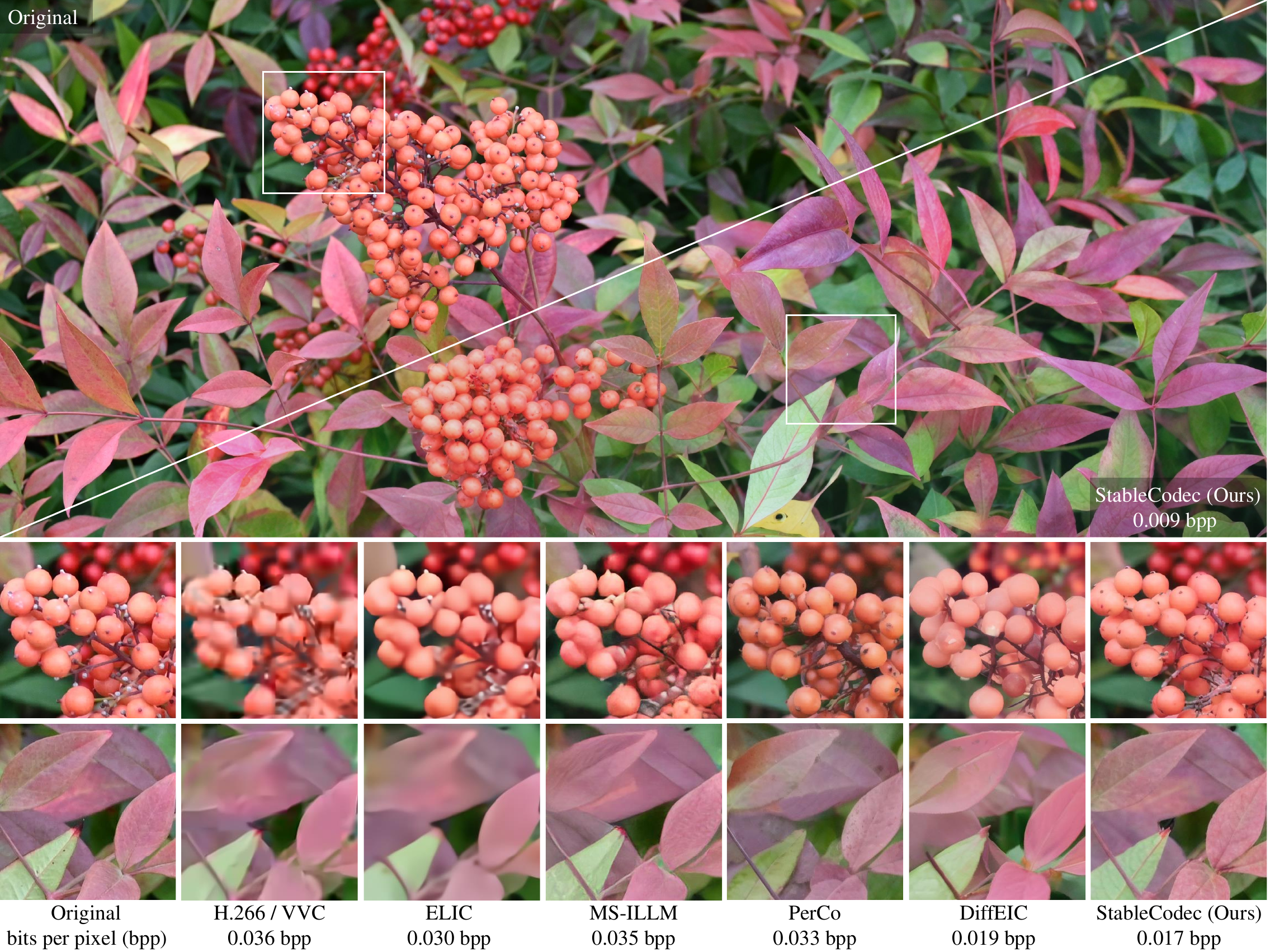}
		\captionof{figure}{\textbf{Visual examples and comparisons when compressing a 4K-resolution image \cite{li2024ustc} at ultra-low bitrates}. The proposed StableCodec produces more realistic and consistent details with fewer bits. In contrast, VVC \cite{bross2021overview}, ELIC \cite{he2022elic} and MS-ILLM \cite{muckley2023improving} reconstructions are blurry, while PerCo \cite{careil2023towards} and DiffEIC \cite{li2024towards} generate inconsistent details against the original images. \textit{Best viewed on screen for details.}}
		\label{teaser}
	\end{center}%
}]



\begin{abstract}
Diffusion-based image compression has shown remarkable potential for achieving ultra-low bitrate coding (less than 0.05 bits per pixel) with high realism, by leveraging the generative priors of large pre-trained text-to-image diffusion models. However, current approaches require a large number of denoising steps at the decoder to generate realistic results under extreme bitrate constraints, limiting their application in real-time compression scenarios. Additionally, these methods often sacrifice reconstruction fidelity, as diffusion models typically fail to guarantee pixel-level consistency. To address these challenges, we introduce \textbf{StableCodec}, which enables one-step diffusion for high-fidelity and high-realism extreme image compression with improved coding efficiency. To achieve ultra-low bitrates, we first develop an efficient Deep Compression Latent Codec to transmit a noisy latent representation for a single-step denoising process. We then propose a Dual-Branch Coding Structure, consisting of a pair of auxiliary encoder and decoder, to enhance reconstruction fidelity. Furthermore, we adopt end-to-end optimization with joint bitrate and pixel-level constraints. Extensive experiments on the CLIC 2020, DIV2K, and Kodak dataset demonstrate that StableCodec outperforms existing methods in terms of FID, KID and DISTS by a significant margin, even at bitrates as low as 0.005 bits per pixel, while maintaining strong fidelity. Additionally, StableCodec achieves inference speeds comparable to mainstream transform coding schemes. All source code are available at \url{https://github.com/LuizScarlet/StableCodec}.
\end{abstract}    
\section{Introduction}
\label{sec:intro}

Image compression is a foundational problem in signal processing. Driven by advances in digital imaging and the widespread use of social platforms, the volume of image data in modern multimedia has grown exponentially, placing increasing demands on the coding efficiency of image compression techniques. Over the past few decades, traditional codecs such as JPEG \cite{wallace1991jpeg} and H.266/VVC \cite{bross2021overview}, along with emerging learning-based methods \cite{balle2016end, balle2018variational, minnen2018joint, cheng2020learned, guo2021causal, he2021checkerboard, jiang2023mlic, liu2023learned, minnen2020channel}, have been widely adopted in real-world image compression applications. However, these methods are typically optimized for rate-distortion performance, and often produce unrealistic and blurry reconstructions, particularly under severe bitrate constraints, as shown in Fig. \ref{teaser}.

To tackle this issue, generative image compression \cite{agustsson2019generative, mentzer2020high} optimized for human perceptual performance has gained increasing attention. These methods are evaluated based on the rate-distortion-perception tradeoff \cite{blau2018perception, blau2019rethinking, yan2021perceptual, yan2022optimally}, and progressively demonstrate their advantages in producing visually appealing reconstructions at lower bitrates compared to traditional codecs or common neural codecs. A prominent research direction \cite{mentzer2020high, agustsson2023multi, he2022po, mentzer2020high, korber2024egic, muckley2023improving} involves integrating a discriminator into the transform coding pipeline \cite{balle2018variational, minnen2020channel, he2022elic}, employing adversarial training to enhance the perceptual quality of reconstructions. Motivated by the impressive generative capabilities, more researchers \cite{li2024towards, xu2024idempotence, yang2024lossy, relic2024lossy, lei2023text+, careil2023towards, hoogeboom2023high, theis2022lossy} have begun exploring the potential of diffusion models, particularly the generative priors in large pre-trained text-to-image (T2I) models, to compensate for severely distorted information at ultra-low bitrates while ensuring perceptually consistent generation. A recent study, PerCo \cite{careil2023towards}, produces realistic results at an extreme bitrate as low as 0.003 bits per pixel (bpp) using a pre-trained latent diffusion model (LDM), highlighting the potential of diffusion-based generative codecs on image compression at more severe bitrates.

\begin{figure}[!t]
    \centering
    \includegraphics[width=0.48\textwidth]{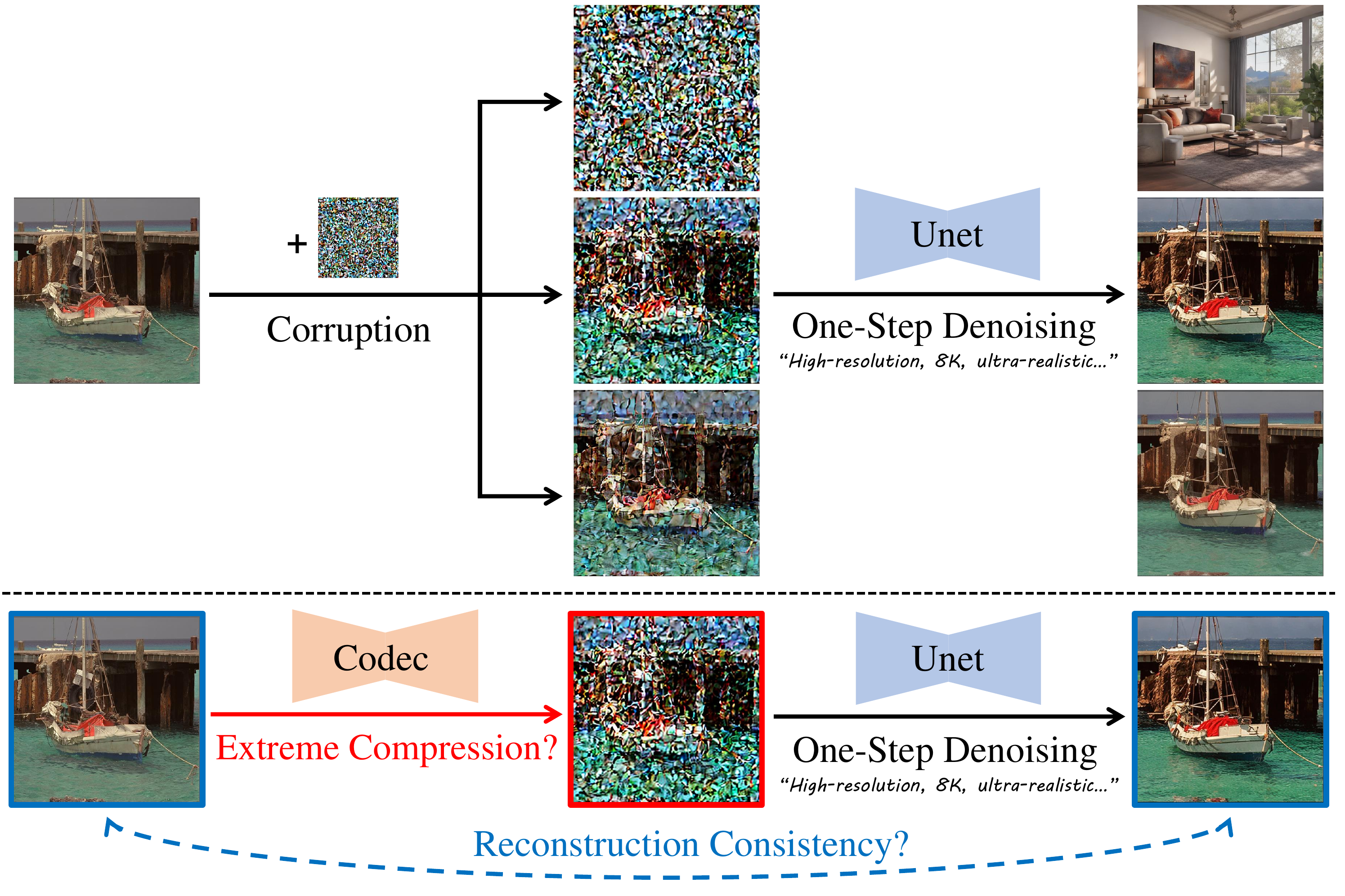}  
    \caption{\textbf{(Top) Illustration of our motivation.} One-step diffusion can produce perceptually consistent results given severely corrupted images and a general prompt. \textbf{(Bottom) Challenges in StableCodec.} How to compress a noisy latent for one-step diffusion using ultra-low bitrates, and how to improve fidelity.}
    \label{fig1}
\end{figure}

Despite these promising advancements, existing methods face two primary limitations inherent to diffusion models. First, they require dozens of denoising steps at the decoder to produce results with sufficient perceptual quality. Second, the reconstructions often deviate from the original images, as diffusion models typically do not guarantee reconstruction consistency. To address the first challenge, we consider leveraging the generative priors in SD-Turbo \cite{sauer2024adversarial}, a distilled version of Stable Diffusion 2.1 \cite{rombach2022high} that enables real-time image synthesis in 1 to 4 denoising steps. Following \cite{zhang2024degradation}, we demonstrate that SD-Turbo can produce perceptually consistent reconstructions with a single-step denoising process, even for severely corrupted inputs and a general positive prompt, as shown in Fig. \ref{fig1}. We thus pose an intuitive question: \textit{Can we compress a noisy latent representation of the original image, which can be effectively denoised in a single-step diffusion process, using an ultra-low bitrate?} Building on these insights, we present \textbf{StableCodec} for extreme image compression, which integrates SD-Turbo with the proposed Deep Compression Latent Codec to compress noisy latents at ultra-low bitrates for a single-step diffusion process.

In response to the second challenge, we introduce a Dual-Branch Coding Structure with a pair of auxiliary encoder and decoder to further enhance reconstruction fidelity. Considering the limitations of the pre-trained VAE encoder on practical entropy coding and reconstruction consistency, we employ a rate-distortion-oriented auxiliary encoder to embed more entropy-aware semantic information for coding decisions. In parallel, we add an auxiliary decoder to perform structure apportionment during the decoding process, improving the generation guidance on the one-step denoising process for more consistent details. To enable end-to-end optimization, we design a two-stage training objective that jointly optimizes bitrate and pixel-level constraints.

Benefit from these designs, StableCodec produces high-fidelity and high-realism reconstructions at ultra-low bitrates as low as 0.005 bpp. Extensive experiments on CLIC 2020 \cite{toderici2020clic}, DIV2K \cite{agustsson2017ntire} and Kodak \cite{kodak} demonstrate that StableCodec sets up a new state-of-the-art performance in terms of FID \cite{heusel2017gans}, KID \cite{binkowski2018demystifying}, and DISTS \cite{ding2020image}, outperforming existing methods by significant margins. In terms of computational complexity, StableCodec offers much faster decoding speeds compared to other diffusion-based competitors, achieving inference times comparable to those of mainstream transform coding schemes. For practical deployment, StableCodec supports inference at arbitrary resolutions with memory consumption less than 9 GB.

We summarize our contributions as follows:
\begin{itemize}
\item We present StableCodec, an extreme image codec integrating one-step diffusion and Deep Compression Latent Codec to achieve ultra-low bitrate compression with high realism, high fidelity and superior coding efficiency.
\item We develop Dual-Branch Coding Structure to improve reconstruction fidelity. A pair of auxiliary encoder and decoder is introduced for semantic enhancement and structure apportionment.
\item StableCodec obtains SOTA FID, KID and DISTS performance on CLIC 2020 and DIV2K dataset, significantly outperforms existing methods at bitrates as low as 0.005 bpp with well-preserved fidelity, and achieves comparable inference speeds with mainstream neural codecs.
\end{itemize}
\section{Related Work}

\subsection{Generative Image Compression}

Learning-based image compression has shown competitive potential compared to traditional standards \cite{wallace1991jpeg, sullivan2012overview, bross2021overview}, leveraging non-linear transforms and joint rate-distortion optimization. \citet{balle2016end} introduced the first end-to-end learned image compression framework, which was subsequently enhanced with the hyperprior \cite{balle2018variational} and context model \cite{minnen2018joint}. Building on this foundation, much work \cite{cheng2020learned, guo2021causal, he2021checkerboard, jiang2023mlic, liu2023learned, qian2022entroformer, minnen2020channel} has been devoted to improving both rate-distortion performance and model practicality.

In practical scenarios, a key challenge is achieving extreme image compression at ultra-low bitrates while maintaining both fidelity and realism \cite{agustsson2019generative}. Traditional image codecs optimized for rate-distortion often produce blurry reconstructions and noticeable artifacts. To address this, \citet{mentzer2020high} introduced HiFiC and the concept of generative image compression, integrating GANs into codec optimization and evaluating performance in terms of the rate-distortion-perception tradeoff \cite{blau2018perception, blau2019rethinking, yan2021perceptual, yan2022optimally}. Subsequent research can be broadly categorized into two main approaches. The first category \cite{agustsson2023multi, he2022po, mentzer2020high, korber2024egic, muckley2023improving} focuses on enhancing transform coding \cite{balle2018variational, minnen2020channel, he2022elic} for human perception by incorporating adversarial losses and optimized discriminator architectures, which typically can be extended to a wide range of bitrate and a flexible decoding control between fidelity and realism \cite{agustsson2023multi, korber2024egic}. The second category \cite{li2024towards, xu2024idempotence, yang2024lossy, relic2024lossy, lei2023text+, careil2023towards, hoogeboom2023high, theis2022lossy} leverages diffusion models for generative image compression. Although these methods show promise for ultra-low bitrate compression, they are often constrained by reconstruction fidelity and inference efficiency due to the multi-step denoising process. Recently, GLC \cite{jia2024generative} introduced transform coding in the generative latent space of VQ-VAE \cite{van2017neural, esser2021taming}, achieving more visually appealing results at ultra-low bitrates.

\subsection{Generative Models and Few-Step Diffusions}

Generative models play a crucial role in image generation. While many architectures, such as VAEs \cite{kingma2013auto} and GANs \cite{goodfellow2020generative}, have been explored, diffusion models \cite{sohl2015deep} have emerged as a powerful alternative, achieving state-of-the-art synthesis quality. Inspired by non-equilibrium statistical physics \cite{sohl2015deep}, diffusion models learn to reverse a noise perturbation process through a Markovian framework. Recent advancements, such as DDPM \cite{ho2020denoising}, DDIM \cite{song2020denoising}, and LDM \cite{rombach2022high}, have significantly reduced computational complexity and improved image synthesis quality, making diffusion-based approaches a dominant force in generative modeling.

To address the inefficiency of iterative denoising, several approaches \cite{salimans2022progressive, song2023consistency, sauer2024adversarial, yin2024improved} aim to reduce the number of denoising steps while maintaining generation quality. These methods train diffusion models to approximate the full denoising trajectory in a single or a few steps, significantly improving inference efficiency. Notably, SD/SDXL-Turbo \cite{sauer2024adversarial} demonstrates image generation in 1 to 4 steps with near-parity quality compared to multi-step models, making real-time diffusion-based applications \cite{parmar2024one, zhang2024degradation, wang2024exploiting} feasible.
\section{Method}

\begin{figure*}[!ht]
    \centering
    \includegraphics[width=\textwidth]{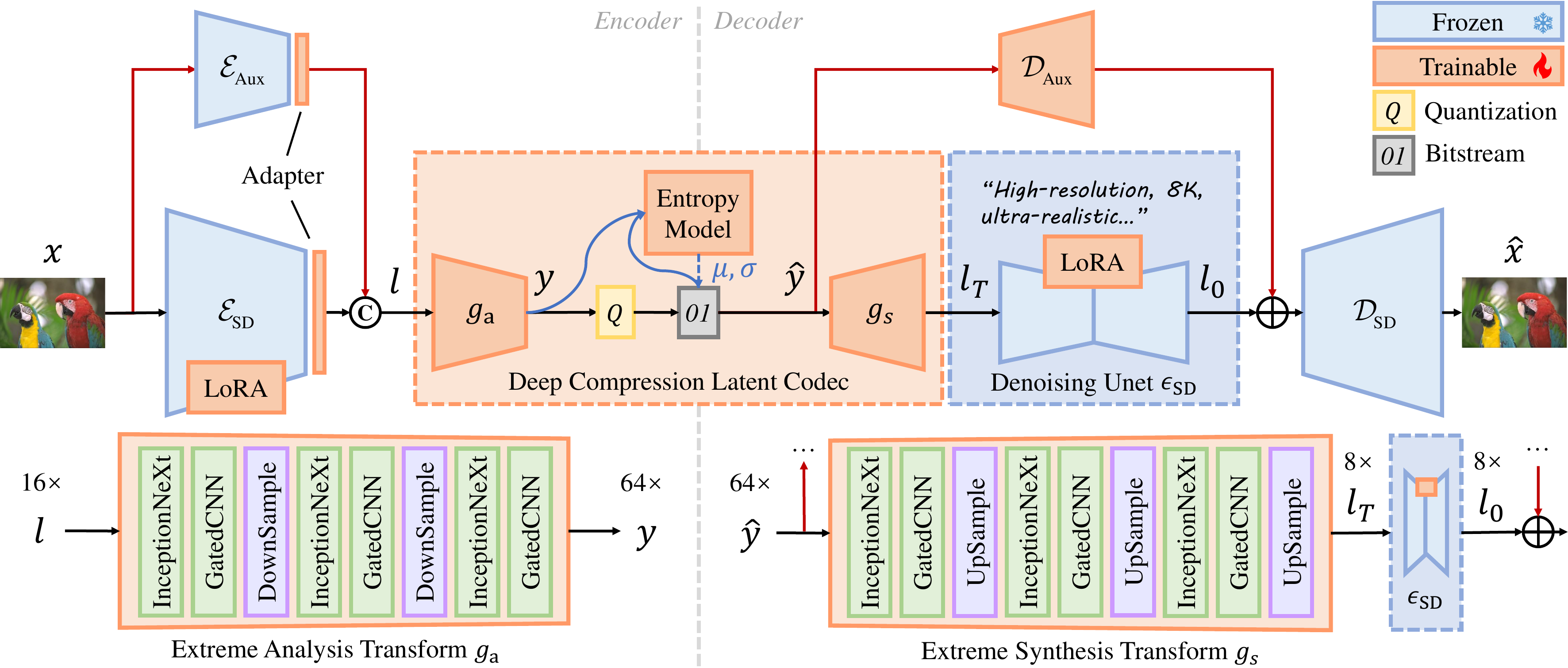}  
    \caption{\textbf{The framework of StableCodec.} We incorporate the proposed Deep Compression Latent Codec to transmit a noisy latent $l_{T}$ for one-step denoising, where $64\times$ denotes a spatial compression ratio of 64. To adjust the latent resolution, we deploy DownSample block and Conv3$\times$3 as adapters after the VAE encoder $\mathcal{E}_{\mathrm{SD}}$ and auxiliary encoder $\mathcal{E}_{\mathrm{Aux}}$, respectively. We use a general prompt in both training and inference. The auxiliary decoder $\mathcal{D}_{\mathrm{Aux}}$ shares a similar structure with $g_{s}$. More details on networks are provided in the supplementary.}
    \label{framework}
\end{figure*}

\subsection{Overview}

In this section, we introduce the overall framework of the proposed StableCodec, built upon SD-Turbo \cite{sauer2024adversarial} with a VAE encoder $\mathcal{E}_{\mathrm{SD}}$, a VAE decoder $\mathcal{D}_{\mathrm{SD}}$ and a denoising Unet $\epsilon_{\mathrm{SD}}$. As shown in Fig. \ref{framework}, we incorporate a Deep Compression Latent Codec to perform extreme transform coding in the VAE latent space. To adapt SD-Turbo for image compression, we integrate LoRA \cite{hu2021lora} into $\mathcal{E}_{\mathrm{SD}}$ and $\epsilon_{\mathrm{SD}}$, while keeping $\mathcal{D}_{\mathrm{SD}}$ unchanged to preserve the generative priors \cite{zhang2024degradation}. Additionally, we introduce a Dual-Branch Coding Structure to enhance reconstruction fidelity, utilizing an auxiliary encoder $\mathcal{E}_{\mathrm{Aux}}$ to embed rich semantic information and an auxiliary decoder $\mathcal{D}_{\mathrm{Aux}}$ to perform structure apportionment. Finally, we optimize StableCodec end-to-end with joint bitrate and pixel-level constraints, achieving high-fidelity and high-realism extreme image compression.

\subsection{Deep Compression Latent Codec}

We design our latent codec using the extreme analysis transform $g_{a}$, extreme synthesis transform $g_{s}$ and a 4-step autoregressive entropy model. To reach ultra-low bitrates, we employ deep compression transform networks for both $g_{a}$ and $g_{s}$. Specifically, $\mathcal{E}_{\mathrm{SD}}$ and $\mathcal{D}_{\mathrm{SD}}$ provide a latent space with a spatial compression ratio of 8 (abbreviated as $8\times$). Unlike mainstream schemes \cite{jia2024generative, li2024towards, he2022elic, jiang2023mlic, liu2023learned, qian2022entroformer, guo2021causal} that perform entropy coding at $16\times$, we further downsample and apply entropy coding for $\hat{y}$ at $64\times$ and the hyperprior \cite{balle2018variational} at $256\times$. Consequently, we use $g_{s}$ to restore the spatial compression ratio to $8\times$ for $\epsilon_{\mathrm{SD}}$ and $\mathcal{D}_{\mathrm{SD}}$. The entire coding process can be formulated as follows:
\begin{align}
  l\ &=\ \mathrm{concat}[\mathcal{E}_{\mathrm{SD}}(x), \mathcal{E}_{\mathrm{Aux}}(x)] \label{eq1}\\
  y\ &=\ g_{a}(l), \ \hat{y}\ =\ Q(y), \ l_{T}\ =\ g_{s}(\hat{y}) \label{eq2}\\
  l_0&=\ [l_{T}-\sqrt{1-\bar{\alpha}_{T}}\cdot\epsilon_{\mathrm{SD}}(l_{T},T)]\ /\ \sqrt{\bar{\alpha}_{T}} \label{eq3}\\
  \hat{x}\ &=\ \mathcal{D}_{\mathrm{SD}}(l_{0}+\mathcal{D}_{\mathrm{Aux}}(\hat{y})) \label{eq4}
\end{align}
In Eq. (\ref{eq1}), we first obtain an intermediate latent $l$ from the input image $x$ through $\mathcal{E}_{\mathrm{SD}}$ and $\mathcal{E}_{\mathrm{Aux}}$. Eq. (\ref{eq2}) is the latent-space transform coding process to produce a noisy latent $l_{T}$ using ultra-low bitrates. Eq. (\ref{eq3}) displays the one-step denoising process with the noise schedule $\left\{\bar{\alpha}_{t}\right\}$ \cite{ho2020denoising} in the $T$-th timestep. Finally, in Eq. (\ref{eq4}), the reconstruction $\hat{x}$ is obtained from $l_{0}$ and $\hat{y}$ using $\mathcal{D}_{\mathrm{SD}}$ and $\mathcal{D}_{\mathrm{Aux}}$. To balance performance and coding latency, we construct efficient $g_{a}$ and $g_{s}$ with InceptionNeXt \cite{yu2024inceptionnext} and GatedCNN \cite{yu2024mambaout}, and build a 4-step antoregressive entropy model with quadtree partition \cite{li2023neural} and latent residual prediction \cite{minnen2020channel}.

\subsection{Dual-Branch Coding Structure}

While deploying the proposed latent codec with LoRA enables image compression with SD-Turbo at ultra-low bitrates, the reconstruction fidelity is limited. In this section, we analyze the reasons and introduce Dual-Branch Coding Structure with a pair of auxiliary encoder and decoder, $\mathcal{E}_{\mathrm{Aux}}$ and $\mathcal{D}_{\mathrm{Aux}}$, to further enhance compression performance.

\subsubsection{Entropy-Aware Semantic Enhancement}

We observed that the VAE in SD-Turbo has several limitations when reconstruction fidelity and practical coding are required. As noted in \cite{rombach2022high} and Table \ref{vae}, this VAE is pre-trained for perceptual compression, which does not preserve pixel-level fidelity as well as a rate-distortion-oriented autoencoder from a typical neural codec \cite{he2022elic}. Additionally, while this VAE provides a compressed representation of the original image, it is still in floating-point format and not optimized for practical entropy coding, making it less suitable for further latent-space ultra-low bitrate compression.

\begin{figure}[!t]
    \centering
    \includegraphics[width=0.48\textwidth]{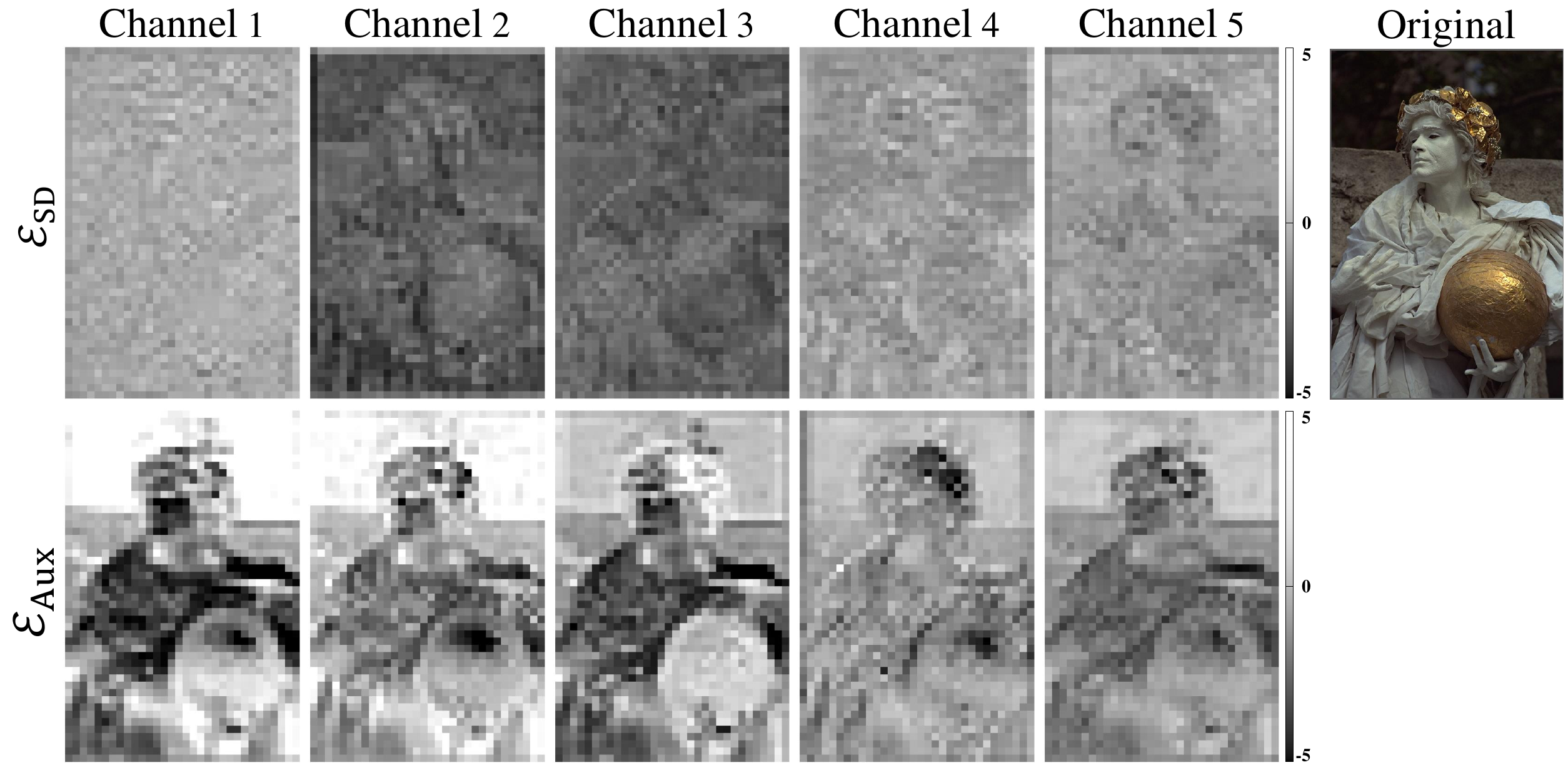}  
    \caption{\textbf{Top-energy channels} learned from different encoders. $\mathcal{E}_{\mathrm{Aux}}$ embeds more pixel-level semantic information into codec.}
    \label{auxe}
\end{figure}

Building on these insights, we introduce the analysis transform of a pre-trained high-bitrate ELIC model \cite{he2022elic} to serve as an auxiliary encoder $\mathcal{E}_{\mathrm{Aux}}$. $\mathcal{E}_{\mathrm{Aux}}$ remains frozen, and provides rich, entropy-aware semantic information of the input images. To combine the latents from different encoders, we introduce trainable adapters that align the latent resolutions to a spatial compression ratio of 16, followed by channel-wise concatenation. During optimization, various pieces of information are learned from different encoders as shown in Fig. \ref{auxe}, where more pixel-level semantic information is embedded into the latent codec through $\mathcal{E}_{\mathrm{Aux}}$.

\begin{table}[!t]
    \centering
    \begin{tabular}{c|cccc}
        \Xhline{1.0pt}
        VAE & PSNR$\uparrow$ & MS-SSIM$\uparrow$ & LPIPS$\downarrow$ & DISTS$\downarrow$ \\
        \hline
        \hline
        SD & 26.65 & 0.9318 & 0.0726 & \textbf{0.0415} \\
        ELIC & \textbf{40.40} & \textbf{0.9961} & \textbf{0.0555} & 0.0707 \\
        \Xhline{1.0pt}
    \end{tabular}
    \caption{\textbf{Reconstruction quality of VAEs} on Kodak \cite{kodak}. The pre-trained VAE in SD performs perceptual compression, while the one in ELIC \cite{he2022elic} preserves more pixel-level information.}
    \label{vae}
\end{table}

\subsubsection{Structure Apportionment}

\begin{figure}[!t]
    \centering
    \includegraphics[width=0.48\textwidth]{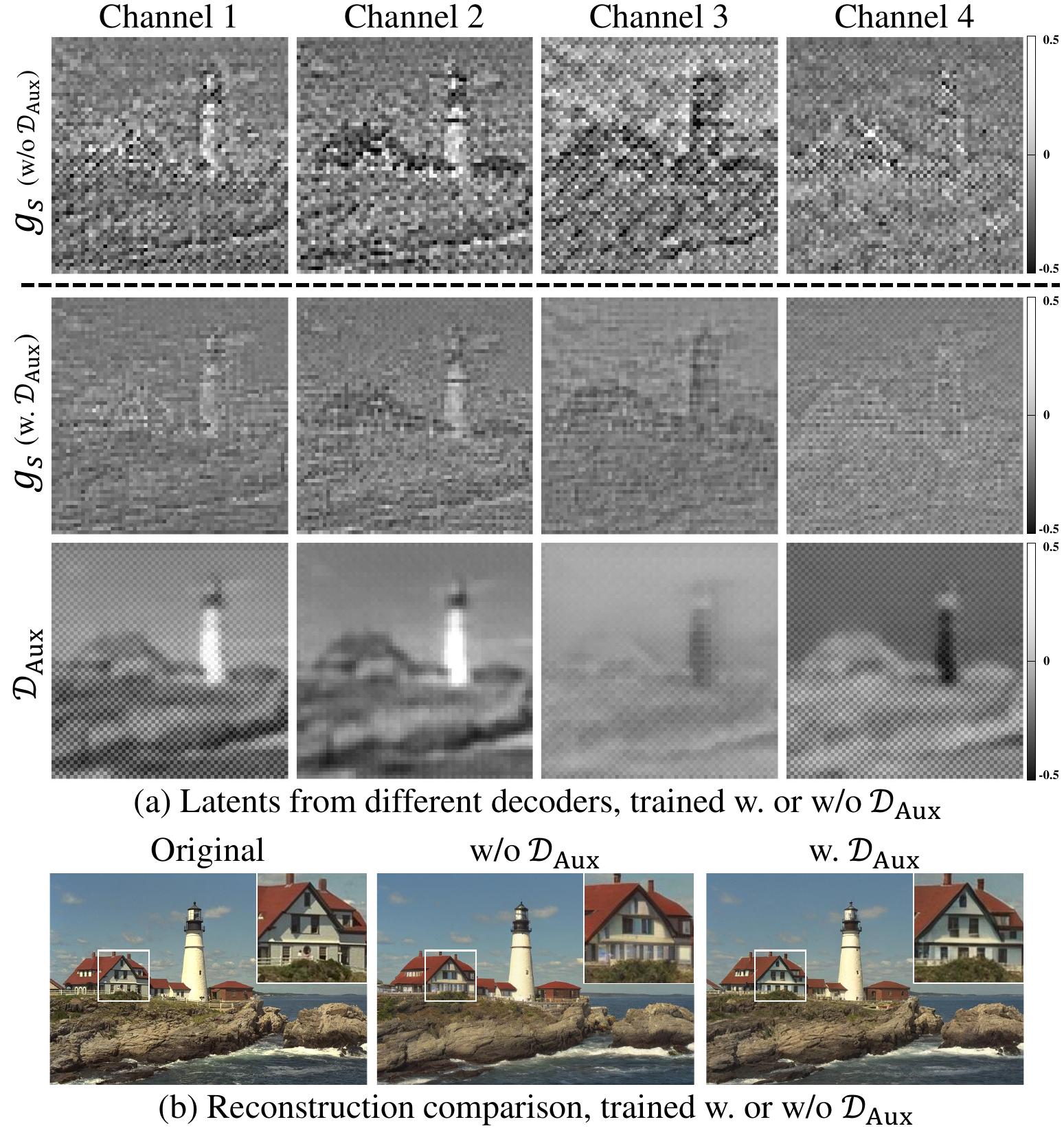}  
    \caption{\textbf{Impact of $\mathcal{D}_{\mathrm{Aux}}$ on latents and reconstructions.}}
    \label{auxd}
\end{figure}

The reconstruction quality of StableCodec heavily depends on how the denoising Unet is conditioned. Since a fixed prompt is used for both training and inference, the one-step denoising process is primarily guided by $l_T$, which is produced by $g_{s}$. This places high demands on the capability of $g_{s}$, resulting in unsatisfactory denoising guidance, as reflected in the reconstructions shown in Fig. \ref{auxd} (b).

To alleviate the decoding burden on $g_{s}$, we introduce an auxiliary decoder $\mathcal{D}_{\mathrm{Aux}}$ to perform an additional decoding branch from $\hat{y}$, bypassing the Unet. This design is motivated by the observation that an extremely compressed bitstream contains mainly the basic structure of images. We distribute and decode these components directly from the bitstream using the auxiliary branch, allowing $g_{s}$ to focus primarily on producing guidance to generate realistic details. Fig. \ref{auxd} compares StableCodec that is trained with or without $\mathcal{D}_{\mathrm{Aux}}$. When trained without $\mathcal{D}_{\mathrm{Aux}}$, $g_{s}$ produces all types of information as it is the only decoding branch. When trained with $\mathcal{D}_{\mathrm{Aux}}$, structural information is routed through $\mathcal{D}_{\mathrm{Aux}}$, while the energy in $g_{s}$ latents drops significantly. Meanwhile, more semantically aligned details are reconstructed in Fig. \ref{auxd} (b), suggesting $g_{s}$ now provides better high-frequency guidance for the denoising Unet $\epsilon_{\mathrm{SD}}$.

\subsection{End-to-End Training Objective}
\label{sec:training}

\begin{figure*}[!h]
    \centering
    \includegraphics[width=\textwidth]{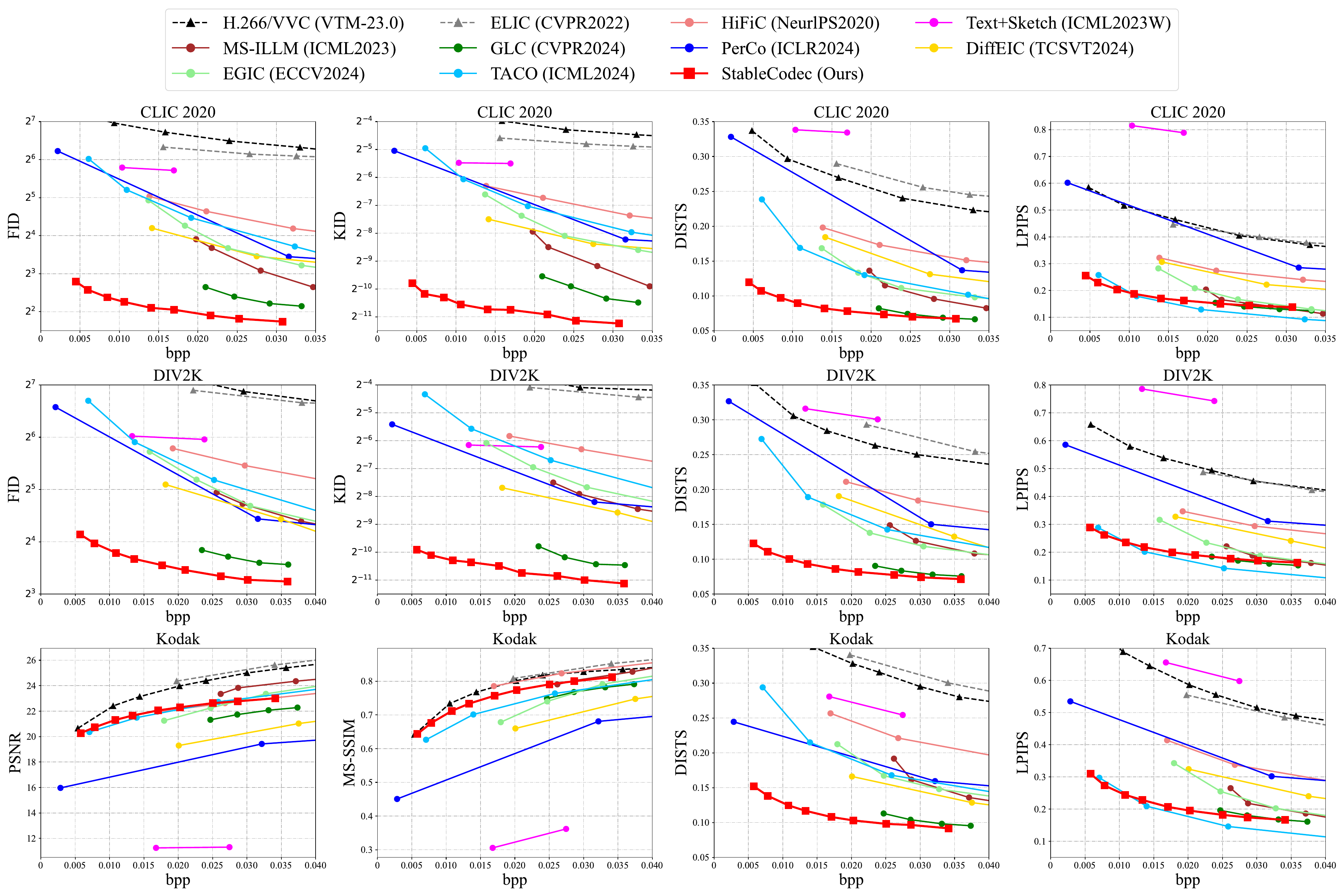}  
    \caption{\textbf{Rate-distortion and rate-perception curve comparisons} of different methods on the CLIC 2020, DIV2K and Kodak dataset.}
    \label{rdpcurve}
\end{figure*}

We adopt end-to-end optimization with joint bitrate and pixel-level restrictions to train StableCodec. Given the original image $x$, the quantized latent $\hat{y}$ and the reconstructed image $\hat{x}$, we construct our training objective based on the standard rate-distortion loss:
\begin{equation}
    \lambda \mathcal{R}(\hat{y}) + \mathcal{D}(x, \hat{x})
\end{equation}
where the bitrate $\mathcal{R}$ and pixel-level distortion $\mathcal{D}$ are balanced by the Lagrange multiplier $\lambda$.

Inspired by \cite{minnen2020channel, he2022elic}, we train StableCodec with a 2-stage implicit bitrate pruning (IBP) strategy. We first train a base model using a smaller $\lambda_{base}$, adapting the latent codec into SD-Turbo under a relaxed bitrate constraint, and warming up with a more expressive transform. In the second stage, we finetune the shared base model with larger $\lambda_{target}$ to reach ultra-low target bitrates. IBP facilitates efficient and stable training, resulting in improved performance.

The distortion term $\mathcal{D}$ includes MSE, LPIPS (using VGG features) \cite{zhang2018unreasonable} and a CLIP \cite{radford2021learning} distance $\mathcal{L}_{CLIP}$, for which we compute the L2-distance between the CLIP embeddings of $x$ and $\hat{x}$. Note that this term is a simplified version from \cite{lee2024neural}, and we find it beneficial for reconstruction at ultra-low bitrates. Additionally, we follow \cite{zhang2024degradation} and incorporate a GAN loss $\mathcal{L}_{adv}$ to narrow the distribution gap between $x$ and $\hat{x}$. We use DINOv2 \cite{oquab2023dinov2} with registers \cite{darcet2023vision} as the discriminator backbone \cite{kumari2022ensembling}. To ensure stable training, we incorporate the GAN only in the second training stage. The full objective can be formulated as:
\begin{align}
    \mathrm{Stage\ \uppercase\expandafter{\romannumeral1}: } &\mathop{\arg\min}\limits_{\theta}\ \lambda_{base}\mathcal{R}(\hat{y}) + \mathcal{D}(x, \hat{x}) \\
    \mathrm{Stage\ \uppercase\expandafter{\romannumeral2}: }  &\mathop{\arg\min}\limits_{\theta}\ \lambda_{target}\mathcal{R}(\hat{y}) + \mathcal{D}(x, \hat{x}) + \beta \mathcal{L}_{adv}
\end{align}
\begin{equation}
\begin{aligned}
  \mathcal{D}(x, \hat{x}) =\  &d_{1}MSE(x, \hat{x}) + d_{2}LPIPS(x, \hat{x}) \\
  +\  &d_{3}\mathcal{L}_{CLIP}(x, \hat{x})
\end{aligned}
\end{equation}
where $\theta$ represents all trainable parameters in StableCodec (Fig. \ref{framework}), $d_{1}$, $d_{2}$, $d_{3}$ and $\beta$ are balancing weights.

\section{Experiments}

\begin{figure*}[!t]
    \centering
    \includegraphics[width=\textwidth]{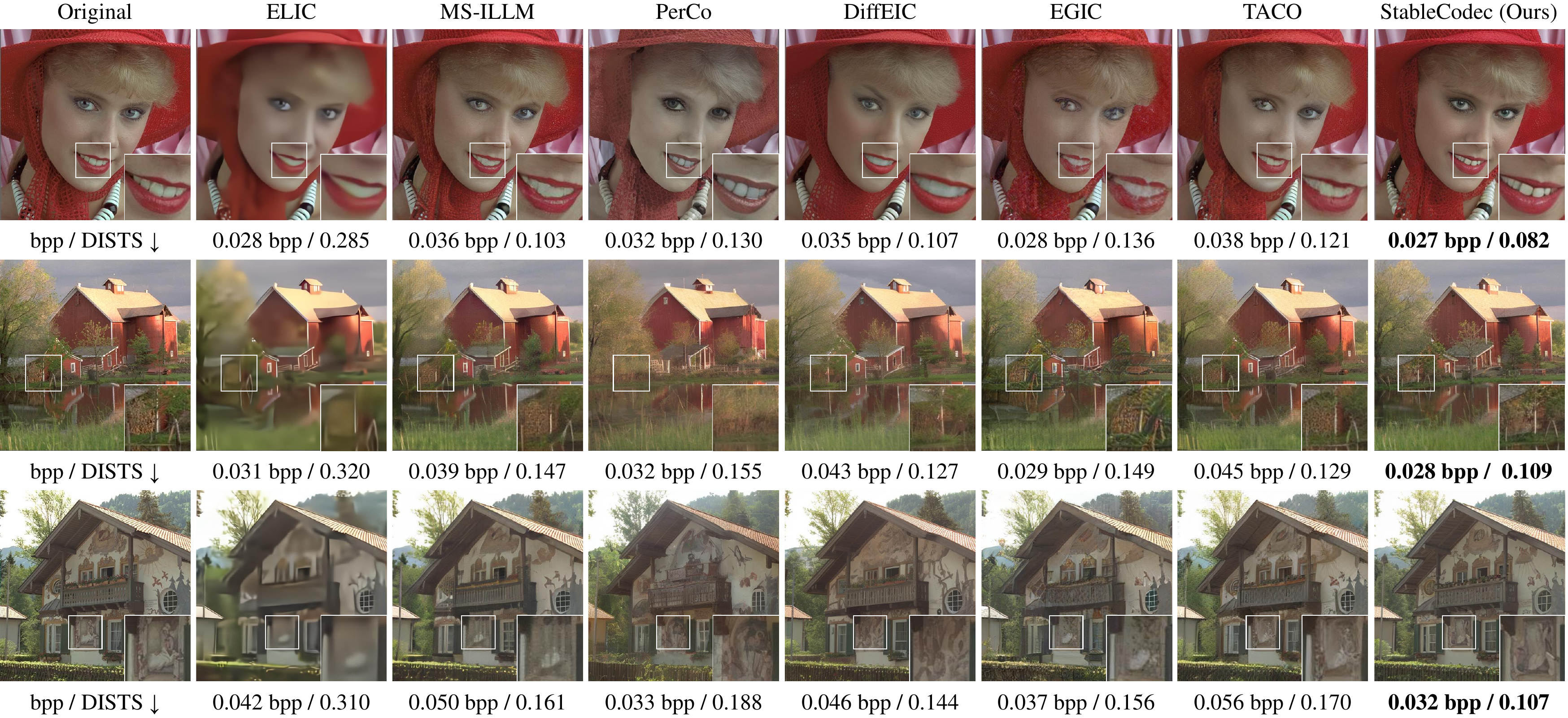}  
    \caption{\textbf{Qualitative comparisons} of different methods on Kodak. \textit{Best viewed on screen for details.}}
    \label{visual}
\end{figure*}

\subsection{Implementation}

\textbf{Training Details.} We use the training set of DF2K \cite{lim2017enhanced} and CLIC 2020 Professional \cite{toderici2020clic} to train StableCodec. During training, we use 512$\times$512 patches with a batch size of 8. The first training stage takes over 100k iterations with a learning rate of $1e^{-4}$ and a $\lambda_{base}$ of $0.5$ (about 0.05bpp). In the second stage, we finetune the base model for another 20k iterations with GAN incorporated and $\lambda_{target} \in \left\{2, 3, 4, 6, 8, 12, 16, 24, 32\right\}$, while the learning rate undergoes $5e^{-5}, 2e^{-5}, 1e^{-5}$ and $1e^{-6}$ for 5k iterations each. We set $d_{1}$, $d_{2}$, $d_{3}$ and $\beta$ to 2, 1, 0.1 and 0.1, repectively. All models are trained using 2 RTX 3090 GPUs. 

\noindent \textbf{Test Data.} We evaluate StableCodec on the test set of CLIC 2020 Professional \cite{toderici2020clic} (CLIC 2020 Test), the validation set of DIV2K \cite{agustsson2017ntire} (DIV2K Val) and Kodak \cite{kodak} following \cite{jia2024generative, li2024towards}. CLIC 2020 Test and DIV2K Val contain 428 and 100 high-quality 2K-resolution natural images, respectively, while Kodak contains 24 natural images with a smaller resolution of 768$\times$512. We evaluate all images with the original resolution as detailed in the supplementary.

\noindent \textbf{Evaluation Metrics.} We employ established metrics to assess the rate-distortion-perception performance of StableCodec. Concretely, we measure bitrate by bits per pixel (bpp), and evaluate perceptual quality using FID \cite{heusel2017gans}, KID \cite{binkowski2018demystifying}, DISTS \cite{ding2020image} and LPIPS \cite{zhang2018unreasonable} (using AlexNet features by default). Meanwhile, we use PSNR and MS-SSIM \cite{wang2003multiscale} to measure the reconstruction fidelity. We follow \cite{muckley2023improving, jia2024generative, li2024towards} to calculate FID and KID on 256$\times$256 patches, and neglect the results on Kodak as it is too small for calculating. Note that pixel-level distortion metrics like LPIPS, PSNR and MS-SSIM have strong limitations when evaluating image compression at ultra-low bitrates \cite{jia2024generative, lei2023text+, ding2020image, careil2023towards}. Therefore, for StableCodec, we focus primarily on FID, KID, and DISTS, which offer a more accurate assessment of quality in severely compressed images. We also provide user study in the supplementary to visually validate the results.

\noindent \textbf{Compared Methods.} We compare StableCodec with advanced image compression methods: \textbf{(1) Traditional Codec} H.266/VVC \cite{bross2021overview} by VTM-23.0 intra coding, \textbf{(2) Neural Codec} ELIC \cite{he2022elic}, \textbf{(3) Generative Codec} HiFiC \cite{mentzer2020high}, Text+Sketch \cite{lei2023text+}, MS-ILLM \cite{muckley2023improving}, PerCo \cite{careil2023towards}, EGIC \cite{korber2024egic}, DiffEIC \cite{li2024towards}, TACO \cite{lee2024neural} and GLC \cite{jia2024generative}. Note that some methods do not release models for ultra-low bitrates, we either re-train or finetune existing weights to reach specific bitrates. For PerCo and GLC that do not have official codes, we use PerCo (SD) \cite{korber2024perco} as a substitute, and request for the results of GLC under the same evaluation approach\footnote{We acknowledge the authors of \cite{jia2024generative} for kindly providing their results.}. We equip PerCo and Text+Sketch with the same inference strategy for a fair comparison on high-resolution images.

\begin{table}[!t]
    \centering
    \setlength{\tabcolsep}{0.9mm}
    \begin{tabular}{c|c|ccc}
        \Xhline{1.0pt}
        Type & Method & \#Steps & Enc. T & Dec. T \\
        \hline
        \hline
        VAE- & ELIC \cite{he2022elic} & - & 0.155 & 0.245 \\
        based & MLIC++ \cite{jiang2023mlicpp} & - & 0.364 & 0.319 \\
        \hline
        GAN- & HiFiC \cite{mentzer2020high} & - & 0.143 & 0.337 \\
        based & MS-ILLM \cite{muckley2023improving} & - & 0.139 & 0.316 \\
        \hline
          & Text+Sketch \cite{lei2023text+} & 25 & 113.252 & 33.560 \\
        Diffusion- & PerCo \cite{careil2023towards} & 20 & 0.287 & 3.742 \\
        based & DiffEIC \cite{li2024towards} & 50 & 0.676 & 7.423 \\
          & StableCodec (Ours) & 1 & 0.159 & 0.326 \\
        \Xhline{1.0pt}
    \end{tabular}
    \caption{\textbf{Encoding and decoding seconds} averaged on Kodak.}
    \label{speed}
\end{table}

\subsection{Main Results}

\subsubsection{Rate-Distortion-Perception Performance}

Fig. \ref{rdpcurve} presents the rate-perception and rate-distortion curves of various methods at ultra-low bitrates over CLIC 2020 Test and DIV2K Val. The proposed StableCodec outperforms all compared methods in terms of FID, KID, and DISTS. Specifically, StableCodec shows a significant improvement over H.266/VVC and ELIC on all perceptual metrics. Compared to generative codec especially previous SOTA GLC \cite{jia2024generative}, StableCodec demonstrates superiority and stability on FID and KID performance with well-preserved fidelity, and reaches extreme bitrates as low as 0.005 bpp. Although TACO \cite{lee2024neural} achieves the best LPIPS performance, it fails to ensure visual quality as FID, KID and DISTS scores are high. The PSNR and MS-SSIM results on CLIC and DIV2K are displayed in the supplementary.

\subsubsection{Qualitative Comparisons}

We provide qualitative results among compared methods in Fig. \ref{visual}. Notably, StableCodec generates more visually-aligned details and realistic textures at ultra-low bitrates, such as the teeth and murals shown in the first and third rows. In contrast, all other methods fail to produce high-realism results with well-preserved fidelity due to the severe bitrate restriction. For example, ELIC and MS-ILLM produce blurry reconstructions, while PerCo and DiffEIC deviate from the original images. EGIC and TACO exhibit noticeable artifacts, particularly in detailed areas.

\subsubsection{Computational Complexity}

We compare the practical complexity of StableCodec with representative schemes in Table \ref{speed} using a single RTX 3090 GPU. Among representative image compression schemes, Diffusion-based methods \cite{careil2023towards, li2024towards, lei2023text+} typically suffer from a much longer decoding time compared to VAE-based \cite{jiang2023mlicpp, he2022elic} or GAN-based \cite{mentzer2020high, muckley2023improving} competitors due to multi-step denoising. In contrast, StableCodec reaches comparable encoding and decoding speed against these methods exploiting one-step denoising, deep compression transforms and efficient entropy model, while achieving significantly better performance at ultra-low bitrates. Detailed runtime analysis is provided in the supplementary. In terms of memory, StableCodec consumes less than 9 GB VRAM with tiling techniques \cite{wang2024exploiting, jimenez2023mixture}, supporting arbitrary-resolution inference on a single GTX 1080Ti GPU.

\section{Ablation Study}

In this section, we conduct ablations to validate the proposed components. For reliable comparison, we compute the BD-rate \cite{bjontegaard2001calculation} with Rate-Y curves on Kodak \cite{kodak} using four target bitrates, where Y denotes specific metrics among PSNR, MS-SSIM, LPIPS, and DISTS.

\begin{table}[!t]
    \centering
    \setlength{\tabcolsep}{0.8mm}
    \begin{tabular}{c|cccc}
        \Xhline{1.0pt}
        Model & \multicolumn{4}{c}{BD-rate$\downarrow$ on the Rate-Y curves} \\
        \cline{2-5}
        Variants & PSNR & MS-SSIM & LPIPS & DISTS \\
        \hline
        \hline
        Base & 0 & 0 & 0 & 0 \\
        + $\mathcal{E}_{\mathrm{Aux}}$ & -20.63\% & -22.05\% & -23.04\% & -28.13\% \\
        + $\mathcal{E}_{\mathrm{Aux}}\ \&\ \mathcal{D}_{\mathrm{Aux}}$ & \textbf{-23.96\%} & \textbf{-28.12\%} & \textbf{-40.66\%} & \textbf{-54.89\%} \\
        \Xhline{1.0pt}
    \end{tabular}
    \caption{\textbf{Ablation study on $\mathcal{E}_{\mathrm{Aux}}$ and $\mathcal{D}_{\mathrm{Aux}}$.}}
    \label{ablation1}
\end{table}

\begin{table}[!t]
    \centering
    \setlength{\tabcolsep}{1.8mm}
    \begin{tabular}{c|cccc}
        \Xhline{1.0pt}
        LoRA & \multicolumn{4}{c}{BD-rate$\downarrow$ on the Rate-Y curves} \\
        \cline{2-5}
        Ranks & PSNR & MS-SSIM & LPIPS & DISTS \\
        \hline
        \hline
        8/8/- & 0 & 0 & 0 & 0 \\
        8/16/- & 1.21\% & -0.35\% & -4.67\% & -5.28\% \\
        16/16/- & -5.62\% & -3.27\% & -6.41\% & -12.78\% \\
        16/32/- & -7.12\% & -5.31\% & \textbf{-13.43\%} & \textbf{-17.96\%} \\
        \hline
        32/32/- & -5.98\% & -5.69\% & -11.21\% & -16.70\% \\
        32/64/- & -5.15\% & -4.98\% & -12.95\% & -17.34\% \\
        \hline
        16/32/4 & \textbf{-21.77\%} & \textbf{-12.28\%} & -8.27\% & -5.13\% \\
        \Xhline{1.0pt}
    \end{tabular}
    \caption{\textbf{Ablation study on LoRA ranks} ($\mathcal{E}_{\mathrm{SD}}$/$\epsilon_{\mathrm{SD}}$/$\mathcal{D}_{\mathrm{SD}}$).}
    \label{ablation2}
\end{table}

\begin{table}[!t]
    \centering
    \setlength{\tabcolsep}{0.65mm}
    \begin{tabular}{ccc|cccc}
        \Xhline{1.0pt}
        \multicolumn{3}{c|}{Training Strategy} & \multicolumn{4}{c}{BD-rate$\downarrow$ on the Rate-Y curves} \\
        \hline
        IBP & $\mathcal{L}_{adv}$ & $\mathcal{L}_{CLIP}$ & PSNR & MS-SSIM & LPIPS & DISTS \\
        \hline
        \hline
        - & - & - & 0 & 0 & 0 & 0 \\
        \checkmark & - & - & \textbf{-24.12\%} & \textbf{-13.67\%} & -21.37\% & -16.60\% \\
        \checkmark & \checkmark & - & 24.41\% & 9.88\% & -36.18\% & -49.24\% \\
        \checkmark & \checkmark & \checkmark & 13.29\% & 5.70\% & \textbf{-38.99\%} & \textbf{-52.95\%} \\
        \Xhline{1.0pt}
    \end{tabular}
    \caption{\textbf{Ablation study on the training strategy components}. Implicit bitrate pruning is abbreviated as IBP. The base one-stage objective only contains bitrate, MSE and LPIPS.}
    \label{ablation3}
\end{table}

\noindent \textbf{Auxiliary Encoder and Decoder.} We begin by providing numerical results for the auxiliary coding branch in Table \ref{ablation1}. Specifically, we construct a base model without $\mathcal{E}_{\mathrm{Aux}}$ and $\mathcal{D}_{\mathrm{Aux}}$, and a variant with $\mathcal{E}_{\mathrm{Aux}}$ only. When $\mathcal{E}_{\mathrm{Aux}}$ is incorporated, more than 20\% bits can be saved to reach the same reconstruction quality. Subsequently, $\mathcal{D}_{\mathrm{Aux}}$ further improves the performance particularly in perceptual quality. As shown in Fig. \ref{auxd}, the purified $g_{s}$ latent provides better guidance for reconstruction consistency.

\noindent \textbf{LoRA Ranks.} We explore the impact of LoRA ranks in Table \ref{ablation2}. Positive results are observed in both distortion and perception as the ranks increase to 16/32, which become our final choice. For larger ranks like 32 and 64, we observe performance degradation as the pre-trained priors may be corrupted. Besides, adding LoRA to the VAE decoder introduces a distortion-perception tradeoff, where PSNR and MS-SSIM improve at the cost of LPIPS and DISTS. To preserve perceptual quality, we leave the decoder unchanged.

\noindent \textbf{Training Strategy.} We perform ablations on our training strategy components in Table \ref{ablation3}. We construct a simplified one-stage objective with bitrate, MSE and LPIPS, then progressively integrate the two-stage implicit bitrate pruning (IBP), adversarial training $\mathcal{L}_{adv}$ and the CLIP distance term $\mathcal{L}_{CLIP}$. We find that IBP improves performance considerably by first adapting the latent codec into the T2I model with relaxed bitrate constraint. Furthermore, incorporating GAN introduces a significant distortion-perception tradeoff since we primarily focus on the perceptual quality. Besides, the CLIP distance alleviates the degradation in distortion and slightly improves perceptual quality.
\section{Conclusion}

In this work, we introduce StableCodec, a novel diffusion-based extreme image compression approach that addresses key limitations of existing methods. By leveraging one-step diffusion in combination with Deep Compression Latent Codec and Dual-Branch Coding Structure, StableCodec achieves ultra-low bitrate compression with high realism, fidelity, and coding efficiency. Extensive experimental evaluations on benchmark datasets demonstrate the superiority of StableCodec in terms of FID, KID, and DISTS, even at extreme bitrates as low as 0.005 bpp, while enabling competitive speeds with mainstream transform coding methods. These results underscore the potential of diffusion models for practical image compression, particularly in real-time coding scenarios where bitrate is severely constrained.

{
    \small
    \bibliographystyle{ieeenat_fullname}
    \bibliography{main}

\begin{thebibliography}{73}
\providecommand{\natexlab}[1]{#1}
\providecommand{\url}[1]{\texttt{#1}}
\expandafter\ifx\csname urlstyle\endcsname\relax
  \providecommand{\doi}[1]{doi: #1}\else
  \providecommand{\doi}{doi: \begingroup \urlstyle{rm}\Url}\fi

\bibitem[til(2023)]{tile}
Tiled diffusion \& vae extension.
\newblock \url{https://github.com/pkuliyi2015/multidiffusion-upscaler-for-automatic1111}, 2023.
\newblock Accessed: 2024-08-27.

\bibitem[Agustsson and Timofte(2017)]{agustsson2017ntire}
Eirikur Agustsson and Radu Timofte.
\newblock Ntire 2017 challenge on single image super-resolution: Dataset and study.
\newblock In \emph{Proceedings of the IEEE conference on computer vision and pattern recognition workshops}, pages 126--135, 2017.

\bibitem[Agustsson et~al.(2019)Agustsson, Tschannen, Mentzer, Timofte, and Gool]{agustsson2019generative}
Eirikur Agustsson, Michael Tschannen, Fabian Mentzer, Radu Timofte, and Luc~Van Gool.
\newblock Generative adversarial networks for extreme learned image compression.
\newblock In \emph{Proceedings of the IEEE/CVF International Conference on Computer Vision}, pages 221--231, 2019.

\bibitem[Agustsson et~al.(2023)Agustsson, Minnen, Toderici, and Mentzer]{agustsson2023multi}
Eirikur Agustsson, David Minnen, George Toderici, and Fabian Mentzer.
\newblock Multi-realism image compression with a conditional generator.
\newblock In \emph{Proceedings of the IEEE/CVF Conference on Computer Vision and Pattern Recognition}, pages 22324--22333, 2023.

\bibitem[Ball{\'e} et~al.(2016)Ball{\'e}, Laparra, and Simoncelli]{balle2016end}
Johannes Ball{\'e}, Valero Laparra, and Eero~P Simoncelli.
\newblock End-to-end optimized image compression.
\newblock \emph{arXiv preprint arXiv:1611.01704}, 2016.

\bibitem[Ball{\'e} et~al.(2018)Ball{\'e}, Minnen, Singh, Hwang, and Johnston]{balle2018variational}
Johannes Ball{\'e}, David Minnen, Saurabh Singh, Sung~Jin Hwang, and Nick Johnston.
\newblock Variational image compression with a scale hyperprior.
\newblock \emph{arXiv preprint arXiv:1802.01436}, 2018.

\bibitem[Bi{\'n}kowski et~al.(2018)Bi{\'n}kowski, Sutherland, Arbel, and Gretton]{binkowski2018demystifying}
Miko{\l}aj Bi{\'n}kowski, Danica~J Sutherland, Michael Arbel, and Arthur Gretton.
\newblock Demystifying mmd gans.
\newblock \emph{arXiv preprint arXiv:1801.01401}, 2018.

\bibitem[Bjontegaard(2001)]{bjontegaard2001calculation}
G Bjontegaard.
\newblock Calculation of average psnr differences between rd-curves.
\newblock \emph{ITU-T SG16 Q}, 6, 2001.

\bibitem[Blau and Michaeli(2018)]{blau2018perception}
Yochai Blau and Tomer Michaeli.
\newblock The perception-distortion tradeoff.
\newblock In \emph{Proceedings of the IEEE conference on computer vision and pattern recognition}, pages 6228--6237, 2018.

\bibitem[Blau and Michaeli(2019)]{blau2019rethinking}
Yochai Blau and Tomer Michaeli.
\newblock Rethinking lossy compression: The rate-distortion-perception tradeoff.
\newblock In \emph{International Conference on Machine Learning}, pages 675--685. PMLR, 2019.

\bibitem[Bross et~al.(2021)Bross, Wang, Ye, Liu, Chen, Sullivan, and Ohm]{bross2021overview}
Benjamin Bross, Ye-Kui Wang, Yan Ye, Shan Liu, Jianle Chen, Gary~J Sullivan, and Jens-Rainer Ohm.
\newblock Overview of the versatile video coding (vvc) standard and its applications.
\newblock \emph{IEEE Transactions on Circuits and Systems for Video Technology}, 31\penalty0 (10):\penalty0 3736--3764, 2021.

\bibitem[Careil et~al.(2023)Careil, Muckley, Verbeek, and Lathuili{\`e}re]{careil2023towards}
Marlene Careil, Matthew~J Muckley, Jakob Verbeek, and St{\'e}phane Lathuili{\`e}re.
\newblock Towards image compression with perfect realism at ultra-low bitrates.
\newblock In \emph{The Twelfth International Conference on Learning Representations}, 2023.

\bibitem[Cheng et~al.(2020)Cheng, Sun, Takeuchi, and Katto]{cheng2020learned}
Zhengxue Cheng, Heming Sun, Masaru Takeuchi, and Jiro Katto.
\newblock Learned image compression with discretized gaussian mixture likelihoods and attention modules.
\newblock In \emph{Proceedings of the IEEE/CVF conference on computer vision and pattern recognition}, pages 7939--7948, 2020.

\bibitem[Choi et~al.(2022)Choi, Lee, Shin, Kim, Kim, and Yoon]{choi2022perception}
Jooyoung Choi, Jungbeom Lee, Chaehun Shin, Sungwon Kim, Hyunwoo Kim, and Sungroh Yoon.
\newblock Perception prioritized training of diffusion models.
\newblock In \emph{Proceedings of the IEEE/CVF Conference on Computer Vision and Pattern Recognition}, pages 11472--11481, 2022.

\bibitem[Darcet et~al.(2023)Darcet, Oquab, Mairal, and Bojanowski]{darcet2023vision}
Timoth{\'e}e Darcet, Maxime Oquab, Julien Mairal, and Piotr Bojanowski.
\newblock Vision transformers need registers.
\newblock \emph{arXiv preprint arXiv:2309.16588}, 2023.

\bibitem[Ding et~al.(2020)Ding, Ma, Wang, and Simoncelli]{ding2020image}
Keyan Ding, Kede Ma, Shiqi Wang, and Eero~P Simoncelli.
\newblock Image quality assessment: Unifying structure and texture similarity.
\newblock \emph{IEEE transactions on pattern analysis and machine intelligence}, 44\penalty0 (5):\penalty0 2567--2581, 2020.

\bibitem[Esser et~al.(2021)Esser, Rombach, and Ommer]{esser2021taming}
Patrick Esser, Robin Rombach, and Bjorn Ommer.
\newblock Taming transformers for high-resolution image synthesis.
\newblock In \emph{Proceedings of the IEEE/CVF conference on computer vision and pattern recognition}, pages 12873--12883, 2021.

\bibitem[Franzen(1993)]{kodak}
Rich Franzen.
\newblock Kodak lossless true color image suite (photocd pcd0992).
\newblock \url{http://r0k.us/graphics/kodak/}, 1993.

\bibitem[Goodfellow et~al.(2020)Goodfellow, Pouget-Abadie, Mirza, Xu, Warde-Farley, Ozair, Courville, and Bengio]{goodfellow2020generative}
Ian Goodfellow, Jean Pouget-Abadie, Mehdi Mirza, Bing Xu, David Warde-Farley, Sherjil Ozair, Aaron Courville, and Yoshua Bengio.
\newblock Generative adversarial networks.
\newblock \emph{Communications of the ACM}, 63\penalty0 (11):\penalty0 139--144, 2020.

\bibitem[Guo et~al.(2021)Guo, Zhang, Feng, and Chen]{guo2021causal}
Zongyu Guo, Zhizheng Zhang, Runsen Feng, and Zhibo Chen.
\newblock Causal contextual prediction for learned image compression.
\newblock \emph{IEEE Transactions on Circuits and Systems for Video Technology}, 32\penalty0 (4):\penalty0 2329--2341, 2021.

\bibitem[He et~al.(2021)He, Zheng, Sun, Wang, and Qin]{he2021checkerboard}
Dailan He, Yaoyan Zheng, Baocheng Sun, Yan Wang, and Hongwei Qin.
\newblock Checkerboard context model for efficient learned image compression.
\newblock In \emph{Proceedings of the IEEE/CVF Conference on Computer Vision and Pattern Recognition}, pages 14771--14780, 2021.

\bibitem[He et~al.(2022{\natexlab{a}})He, Yang, Peng, Ma, Qin, and Wang]{he2022elic}
Dailan He, Ziming Yang, Weikun Peng, Rui Ma, Hongwei Qin, and Yan Wang.
\newblock Elic: Efficient learned image compression with unevenly grouped space-channel contextual adaptive coding.
\newblock In \emph{Proceedings of the IEEE/CVF Conference on Computer Vision and Pattern Recognition}, pages 5718--5727, 2022{\natexlab{a}}.

\bibitem[He et~al.(2022{\natexlab{b}})He, Yang, Yu, Xu, Luo, Chen, Gao, Shi, Qin, and Wang]{he2022po}
Dailan He, Ziming Yang, Hongjiu Yu, Tongda Xu, Jixiang Luo, Yuan Chen, Chenjian Gao, Xinjie Shi, Hongwei Qin, and Yan Wang.
\newblock Po-elic: Perception-oriented efficient learned image coding.
\newblock In \emph{Proceedings of the IEEE/CVF Conference on Computer Vision and Pattern Recognition}, pages 1764--1769, 2022{\natexlab{b}}.

\bibitem[Heusel et~al.(2017)Heusel, Ramsauer, Unterthiner, Nessler, and Hochreiter]{heusel2017gans}
Martin Heusel, Hubert Ramsauer, Thomas Unterthiner, Bernhard Nessler, and Sepp Hochreiter.
\newblock Gans trained by a two time-scale update rule converge to a local nash equilibrium.
\newblock \emph{Advances in neural information processing systems}, 30, 2017.

\bibitem[Ho et~al.(2020)Ho, Jain, and Abbeel]{ho2020denoising}
Jonathan Ho, Ajay Jain, and Pieter Abbeel.
\newblock Denoising diffusion probabilistic models.
\newblock \emph{Advances in neural information processing systems}, 33:\penalty0 6840--6851, 2020.

\bibitem[Hoogeboom et~al.(2023)Hoogeboom, Agustsson, Mentzer, Versari, Toderici, and Theis]{hoogeboom2023high}
Emiel Hoogeboom, Eirikur Agustsson, Fabian Mentzer, Luca Versari, George Toderici, and Lucas Theis.
\newblock High-fidelity image compression with score-based generative models.
\newblock \emph{arXiv preprint arXiv:2305.18231}, 2023.

\bibitem[Hu et~al.(2021)Hu, Shen, Wallis, Allen-Zhu, Li, Wang, Wang, and Chen]{hu2021lora}
Edward~J Hu, Yelong Shen, Phillip Wallis, Zeyuan Allen-Zhu, Yuanzhi Li, Shean Wang, Lu Wang, and Weizhu Chen.
\newblock Lora: Low-rank adaptation of large language models.
\newblock \emph{arXiv preprint arXiv:2106.09685}, 2021.

\bibitem[Jia et~al.(2024)Jia, Li, Li, Li, and Lu]{jia2024generative}
Zhaoyang Jia, Jiahao Li, Bin Li, Houqiang Li, and Yan Lu.
\newblock Generative latent coding for ultra-low bitrate image compression.
\newblock In \emph{Proceedings of the IEEE/CVF Conference on Computer Vision and Pattern Recognition}, pages 26088--26098, 2024.

\bibitem[Jiang and Wang(2023)]{jiang2023mlicpp}
Wei Jiang and Ronggang Wang.
\newblock Mlic++: Linear complexity multi-reference entropy modeling for learned image compression.
\newblock In \emph{ICML 2023 Workshop Neural Compression: From Information Theory to Applications}, 2023.

\bibitem[Jiang et~al.(2023)Jiang, Yang, Zhai, Ning, Gao, and Wang]{jiang2023mlic}
Wei Jiang, Jiayu Yang, Yongqi Zhai, Peirong Ning, Feng Gao, and Ronggang Wang.
\newblock Mlic: Multi-reference entropy model for learned image compression.
\newblock In \emph{Proceedings of the 31st ACM International Conference on Multimedia}, pages 7618--7627, 2023.

\bibitem[Jim{\'e}nez(2023)]{jimenez2023mixture}
{\'A}lvaro~Barbero Jim{\'e}nez.
\newblock Mixture of diffusers for scene composition and high resolution image generation.
\newblock \emph{arXiv preprint arXiv:2302.02412}, 2023.

\bibitem[Kingma(2013)]{kingma2013auto}
Diederik~P Kingma.
\newblock Auto-encoding variational bayes.
\newblock \emph{arXiv preprint arXiv:1312.6114}, 2013.

\bibitem[K{\"o}rber et~al.(2024{\natexlab{a}})K{\"o}rber, Kromer, Siebert, Hauke, Mueller-Gritschneder, and Schuller]{korber2024egic}
Nikolai K{\"o}rber, Eduard Kromer, Andreas Siebert, Sascha Hauke, Daniel Mueller-Gritschneder, and Bj{\"o}rn Schuller.
\newblock Egic: enhanced low-bit-rate generative image compression guided by semantic segmentation.
\newblock In \emph{European Conference on Computer Vision}, pages 202--220. Springer, 2024{\natexlab{a}}.

\bibitem[K{\"o}rber et~al.(2024{\natexlab{b}})K{\"o}rber, Kromer, Siebert, Hauke, Mueller-Gritschneder, and Schuller]{korber2024perco}
Nikolai K{\"o}rber, Eduard Kromer, Andreas Siebert, Sascha Hauke, Daniel Mueller-Gritschneder, and Bj{\"o}rn Schuller.
\newblock Perco (sd): Open perceptual compression.
\newblock \emph{arXiv preprint arXiv:2409.20255}, 2024{\natexlab{b}}.

\bibitem[Kumari et~al.(2022)Kumari, Zhang, Shechtman, and Zhu]{kumari2022ensembling}
Nupur Kumari, Richard Zhang, Eli Shechtman, and Jun-Yan Zhu.
\newblock Ensembling off-the-shelf models for gan training.
\newblock In \emph{Proceedings of the IEEE/CVF conference on computer vision and pattern recognition}, pages 10651--10662, 2022.

\bibitem[Lee et~al.(2024)Lee, Kim, Kim, Kim, Oh, and Lee]{lee2024neural}
Hagyeong Lee, Minkyu Kim, Jun-Hyuk Kim, Seungeon Kim, Dokwan Oh, and Jaeho Lee.
\newblock Neural image compression with text-guided encoding for both pixel-level and perceptual fidelity.
\newblock \emph{arXiv preprint arXiv:2403.02944}, 2024.

\bibitem[Lei et~al.(2023)Lei, Uslu, Hassani, and Bidokhti]{lei2023text+}
Eric Lei, Yi{\u{g}}it~Berkay Uslu, Hamed Hassani, and Shirin~Saeedi Bidokhti.
\newblock Text+ sketch: Image compression at ultra low rates.
\newblock \emph{arXiv preprint arXiv:2307.01944}, 2023.

\bibitem[Li et~al.(2023)Li, Li, and Lu]{li2023neural}
Jiahao Li, Bin Li, and Yan Lu.
\newblock Neural video compression with diverse contexts.
\newblock In \emph{Proceedings of the IEEE/CVF Conference on Computer Vision and Pattern Recognition}, pages 22616--22626, 2023.

\bibitem[Li et~al.(2024{\natexlab{a}})Li, Liao, Tang, Zhang, Li, Bian, Sheng, Feng, Li, Gao, et~al.]{li2024ustc}
Zhuoyuan Li, Junqi Liao, Chuanbo Tang, Haotian Zhang, Yuqi Li, Yifan Bian, Xihua Sheng, Xinmin Feng, Yao Li, Changsheng Gao, et~al.
\newblock Ustc-td: A test dataset and benchmark for image and video coding in 2020s.
\newblock \emph{arXiv preprint arXiv:2409.08481}, 2024{\natexlab{a}}.

\bibitem[Li et~al.(2024{\natexlab{b}})Li, Zhou, Wei, Ge, and Jiang]{li2024towards}
Zhiyuan Li, Yanhui Zhou, Hao Wei, Chenyang Ge, and Jingwen Jiang.
\newblock Towards extreme image compression with latent feature guidance and diffusion prior.
\newblock \emph{IEEE Transactions on Circuits and Systems for Video Technology}, 2024{\natexlab{b}}.

\bibitem[Lim et~al.(2017)Lim, Son, Kim, Nah, and Mu~Lee]{lim2017enhanced}
Bee Lim, Sanghyun Son, Heewon Kim, Seungjun Nah, and Kyoung Mu~Lee.
\newblock Enhanced deep residual networks for single image super-resolution.
\newblock In \emph{Proceedings of the IEEE conference on computer vision and pattern recognition workshops}, pages 136--144, 2017.

\bibitem[Liu et~al.(2023)Liu, Sun, and Katto]{liu2023learned}
Jinming Liu, Heming Sun, and Jiro Katto.
\newblock Learned image compression with mixed transformer-cnn architectures.
\newblock In \emph{Proceedings of the IEEE/CVF Conference on Computer Vision and Pattern Recognition}, pages 14388--14397, 2023.

\bibitem[Mentzer et~al.(2020)Mentzer, Toderici, Tschannen, and Agustsson]{mentzer2020high}
Fabian Mentzer, George~D Toderici, Michael Tschannen, and Eirikur Agustsson.
\newblock High-fidelity generative image compression.
\newblock \emph{Advances in Neural Information Processing Systems}, 33:\penalty0 11913--11924, 2020.

\bibitem[Minnen and Singh(2020)]{minnen2020channel}
David Minnen and Saurabh Singh.
\newblock Channel-wise autoregressive entropy models for learned image compression.
\newblock In \emph{2020 IEEE International Conference on Image Processing (ICIP)}, pages 3339--3343. IEEE, 2020.

\bibitem[Minnen et~al.(2018)Minnen, Ball{\'e}, and Toderici]{minnen2018joint}
David Minnen, Johannes Ball{\'e}, and George~D Toderici.
\newblock Joint autoregressive and hierarchical priors for learned image compression.
\newblock \emph{Advances in neural information processing systems}, 31, 2018.

\bibitem[Muckley et~al.(2023)Muckley, El-Nouby, Ullrich, J{\'e}gou, and Verbeek]{muckley2023improving}
Matthew~J Muckley, Alaaeldin El-Nouby, Karen Ullrich, Herv{\'e} J{\'e}gou, and Jakob Verbeek.
\newblock Improving statistical fidelity for neural image compression with implicit local likelihood models.
\newblock In \emph{International Conference on Machine Learning}, pages 25426--25443. PMLR, 2023.

\bibitem[Oquab et~al.(2023)Oquab, Darcet, Moutakanni, Vo, Szafraniec, Khalidov, Fernandez, Haziza, Massa, El-Nouby, et~al.]{oquab2023dinov2}
Maxime Oquab, Timoth{\'e}e Darcet, Th{\'e}o Moutakanni, Huy Vo, Marc Szafraniec, Vasil Khalidov, Pierre Fernandez, Daniel Haziza, Francisco Massa, Alaaeldin El-Nouby, et~al.
\newblock Dinov2: Learning robust visual features without supervision.
\newblock \emph{arXiv preprint arXiv:2304.07193}, 2023.

\bibitem[Parmar et~al.(2024)Parmar, Park, Narasimhan, and Zhu]{parmar2024one}
Gaurav Parmar, Taesung Park, Srinivasa Narasimhan, and Jun-Yan Zhu.
\newblock One-step image translation with text-to-image models.
\newblock \emph{arXiv preprint arXiv:2403.12036}, 2024.

\bibitem[Qian et~al.(2022)Qian, Lin, Sun, Tan, and Jin]{qian2022entroformer}
Yichen Qian, Ming Lin, Xiuyu Sun, Zhiyu Tan, and Rong Jin.
\newblock Entroformer: A transformer-based entropy model for learned image compression.
\newblock \emph{arXiv preprint arXiv:2202.05492}, 2022.

\bibitem[Radford et~al.(2021)Radford, Kim, Hallacy, Ramesh, Goh, Agarwal, Sastry, Askell, Mishkin, Clark, et~al.]{radford2021learning}
Alec Radford, Jong~Wook Kim, Chris Hallacy, Aditya Ramesh, Gabriel Goh, Sandhini Agarwal, Girish Sastry, Amanda Askell, Pamela Mishkin, Jack Clark, et~al.
\newblock Learning transferable visual models from natural language supervision.
\newblock In \emph{International conference on machine learning}, pages 8748--8763. PMLR, 2021.

\bibitem[Relic et~al.(2024)Relic, Azevedo, Gross, and Schroers]{relic2024lossy}
Lucas Relic, Roberto Azevedo, Markus Gross, and Christopher Schroers.
\newblock Lossy image compression with foundation diffusion models.
\newblock In \emph{European Conference on Computer Vision}, pages 303--319. Springer, 2024.

\bibitem[Rombach et~al.(2022)Rombach, Blattmann, Lorenz, Esser, and Ommer]{rombach2022high}
Robin Rombach, Andreas Blattmann, Dominik Lorenz, Patrick Esser, and Bj{\"o}rn Ommer.
\newblock High-resolution image synthesis with latent diffusion models.
\newblock In \emph{Proceedings of the IEEE/CVF conference on computer vision and pattern recognition}, pages 10684--10695, 2022.

\bibitem[Salimans and Ho(2022)]{salimans2022progressive}
Tim Salimans and Jonathan Ho.
\newblock Progressive distillation for fast sampling of diffusion models.
\newblock \emph{arXiv preprint arXiv:2202.00512}, 2022.

\bibitem[Sauer et~al.(2024)Sauer, Lorenz, Blattmann, and Rombach]{sauer2024adversarial}
Axel Sauer, Dominik Lorenz, Andreas Blattmann, and Robin Rombach.
\newblock Adversarial diffusion distillation.
\newblock In \emph{European Conference on Computer Vision}, pages 87--103. Springer, 2024.

\bibitem[Sohl-Dickstein et~al.(2015)Sohl-Dickstein, Weiss, Maheswaranathan, and Ganguli]{sohl2015deep}
Jascha Sohl-Dickstein, Eric Weiss, Niru Maheswaranathan, and Surya Ganguli.
\newblock Deep unsupervised learning using nonequilibrium thermodynamics.
\newblock In \emph{International conference on machine learning}, pages 2256--2265. PMLR, 2015.

\bibitem[Song et~al.(2020)Song, Meng, and Ermon]{song2020denoising}
Jiaming Song, Chenlin Meng, and Stefano Ermon.
\newblock Denoising diffusion implicit models.
\newblock \emph{arXiv preprint arXiv:2010.02502}, 2020.

\bibitem[Song et~al.(2023)Song, Dhariwal, Chen, and Sutskever]{song2023consistency}
Yang Song, Prafulla Dhariwal, Mark Chen, and Ilya Sutskever.
\newblock Consistency models.
\newblock \emph{arXiv preprint arXiv:2303.01469}, 2023.

\bibitem[Sullivan et~al.(2012)Sullivan, Ohm, Han, and Wiegand]{sullivan2012overview}
Gary~J Sullivan, Jens-Rainer Ohm, Woo-Jin Han, and Thomas Wiegand.
\newblock Overview of the high efficiency video coding (hevc) standard.
\newblock \emph{IEEE Transactions on circuits and systems for video technology}, 22\penalty0 (12):\penalty0 1649--1668, 2012.

\bibitem[Theis et~al.(2022)Theis, Salimans, Hoffman, and Mentzer]{theis2022lossy}
Lucas Theis, Tim Salimans, Matthew~D Hoffman, and Fabian Mentzer.
\newblock Lossy compression with gaussian diffusion.
\newblock \emph{arXiv preprint arXiv:2206.08889}, 2022.

\bibitem[Toderici et~al.(2020)Toderici, Theis, Johnston, Agustsson, Mentzer, Ball{\'e}, Shi, and Timofte]{toderici2020clic}
George Toderici, Lucas Theis, Nick Johnston, Eirikur Agustsson, Fabian Mentzer, Johannes Ball{\'e}, Wenzhe Shi, and Radu Timofte.
\newblock Clic 2020: Challenge on learned image compression, 2020, 2020.

\bibitem[Van Den~Oord et~al.(2017)Van Den~Oord, Vinyals, et~al.]{van2017neural}
Aaron Van Den~Oord, Oriol Vinyals, et~al.
\newblock Neural discrete representation learning.
\newblock \emph{Advances in neural information processing systems}, 30, 2017.

\bibitem[Wallace(1991)]{wallace1991jpeg}
Gregory~K Wallace.
\newblock The jpeg still picture compression standard.
\newblock \emph{Communications of the ACM}, 34\penalty0 (4):\penalty0 30--44, 1991.

\bibitem[Wang et~al.(2024)Wang, Yue, Zhou, Chan, and Loy]{wang2024exploiting}
Jianyi Wang, Zongsheng Yue, Shangchen Zhou, Kelvin~CK Chan, and Chen~Change Loy.
\newblock Exploiting diffusion prior for real-world image super-resolution.
\newblock \emph{International Journal of Computer Vision}, pages 1--21, 2024.

\bibitem[Wang et~al.(2003)Wang, Simoncelli, and Bovik]{wang2003multiscale}
Zhou Wang, Eero~P Simoncelli, and Alan~C Bovik.
\newblock Multiscale structural similarity for image quality assessment.
\newblock In \emph{The Thrity-Seventh Asilomar Conference on Signals, Systems \& Computers, 2003}, pages 1398--1402. Ieee, 2003.

\bibitem[Xu et~al.(2024)Xu, Zhu, He, Li, Guo, Wang, Wang, Qin, Wang, Liu, et~al.]{xu2024idempotence}
Tongda Xu, Ziran Zhu, Dailan He, Yanghao Li, Lina Guo, Yuanyuan Wang, Zhe Wang, Hongwei Qin, Yan Wang, Jingjing Liu, et~al.
\newblock Idempotence and perceptual image compression.
\newblock \emph{arXiv preprint arXiv:2401.08920}, 2024.

\bibitem[Yan et~al.(2021)Yan, Wen, Ying, Ma, and Liu]{yan2021perceptual}
Zeyu Yan, Fei Wen, Rendong Ying, Chao Ma, and Peilin Liu.
\newblock On perceptual lossy compression: The cost of perceptual reconstruction and an optimal training framework.
\newblock In \emph{International Conference on Machine Learning}, pages 11682--11692. PMLR, 2021.

\bibitem[Yan et~al.(2022)Yan, Wen, and Liu]{yan2022optimally}
Zeyu Yan, Fei Wen, and Peilin Liu.
\newblock Optimally controllable perceptual lossy compression.
\newblock \emph{arXiv preprint arXiv:2206.10082}, 2022.

\bibitem[Yang and Mandt(2024)]{yang2024lossy}
Ruihan Yang and Stephan Mandt.
\newblock Lossy image compression with conditional diffusion models.
\newblock \emph{Advances in Neural Information Processing Systems}, 36, 2024.

\bibitem[Yin et~al.(2024)Yin, Gharbi, Park, Zhang, Shechtman, Durand, and Freeman]{yin2024improved}
Tianwei Yin, Micha{\"e}l Gharbi, Taesung Park, Richard Zhang, Eli Shechtman, Fredo Durand, and William~T Freeman.
\newblock Improved distribution matching distillation for fast image synthesis.
\newblock \emph{arXiv preprint arXiv:2405.14867}, 2024.

\bibitem[Yu and Wang(2024)]{yu2024mambaout}
Weihao Yu and Xinchao Wang.
\newblock Mambaout: Do we really need mamba for vision?
\newblock \emph{arXiv preprint arXiv:2405.07992}, 2024.

\bibitem[Yu et~al.(2024)Yu, Zhou, Yan, and Wang]{yu2024inceptionnext}
Weihao Yu, Pan Zhou, Shuicheng Yan, and Xinchao Wang.
\newblock Inceptionnext: When inception meets convnext.
\newblock In \emph{Proceedings of the IEEE/CVF Conference on Computer Vision and Pattern Recognition}, pages 5672--5683, 2024.

\bibitem[Zhang et~al.(2024)Zhang, Yue, Pei, Ren, and Cao]{zhang2024degradation}
Aiping Zhang, Zongsheng Yue, Renjing Pei, Wenqi Ren, and Xiaochun Cao.
\newblock Degradation-guided one-step image super-resolution with diffusion priors.
\newblock \emph{arXiv preprint arXiv:2409.17058}, 2024.

\bibitem[Zhang et~al.(2018)Zhang, Isola, Efros, Shechtman, and Wang]{zhang2018unreasonable}
Richard Zhang, Phillip Isola, Alexei~A Efros, Eli Shechtman, and Oliver Wang.
\newblock The unreasonable effectiveness of deep features as a perceptual metric.
\newblock In \emph{Proceedings of the IEEE conference on computer vision and pattern recognition}, pages 586--595, 2018.

\end{thebibliography}
}
\clearpage
\appendix

\clearpage

\setcounter{figure}{7}
\setcounter{table}{5}
\setcounter{equation}{8}

\maketitlesupplementary

\renewcommand\thesection{\Alph{section}}

\section{Inference for Arbitrary Resolution}

Diffusion models typically face scalability issues when dealing with high-resolution images, often yielding inferior results while incurring significantly increased computational costs. Consequently, existing diffusion-based codecs \cite{lei2023text+, careil2023towards, li2024towards} primarily target small images with resolutions around 512×512 or resized images. To enhance the practicality of StableCodec, we adopt a tiled VAE approach \cite{tile} to split high-resolution images into tiles and process them sequentially in both the VAE encoder and decoder. For one-step denoising, we employ a similar latent aggregation strategy \cite{jimenez2023mixture, wang2024exploiting}, which processes latent patches individually and aggregates overlapping pixels using a Gaussian weight map. These methods enable StableCodec to support arbitrary-resolution inference with memory consumption under 9 GB, greatly improving its efficiency and practicality for real-world deployment.

\begin{figure}[!t]
    \centering
    \includegraphics[width=0.47\textwidth]{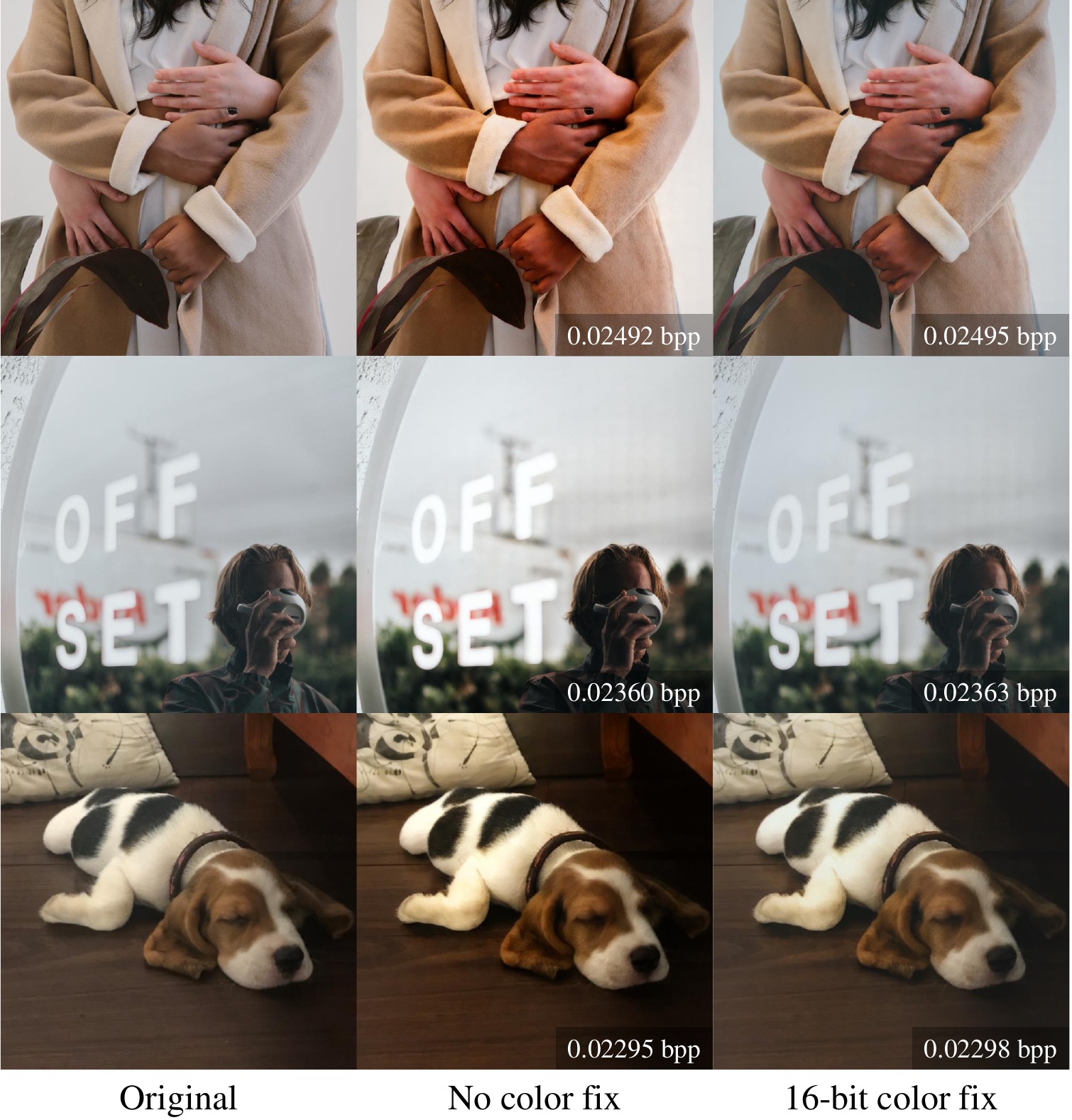}  
    \caption{\textbf{Visual examples of color fix} from CLIC 2020 \cite{toderici2020clic}. 16-bit color fix brings clear refinement with negligible bits increase.}
    \label{color}
\end{figure}

However, we observe that StableCodec sometimes produces color shifts when reconstructing high-resolution images, as illustrated in Fig. \ref{color}. This issue has also been noted in \cite{wang2024exploiting, choi2022perception}. To address this, we apply a quantized version of adaptive instance normalization \cite{wang2024exploiting} on the reconstructed high-resolution image $\hat{x}$, aligning its mean ($\mu_{\hat{x}}$) and variance ($\sigma_{\hat{x}}$) with those of the original image ($\mu_{x}$ and $\sigma_{x}$):
\begin{equation}
    \hat{x}^{c} = \frac{\hat{x} - \mu_{\hat{x}}}{\sigma_{\hat{x}}} \cdot \hat{\sigma_{x}} + \hat{\mu_{x}}
\end{equation}
where $\hat{\mu_{x}}$ and $\hat{\sigma_{x}}$ are 16-bit-quantized from $\mu_{x}$ and $\sigma_{x}$:
\begin{align}
  \hat{\mu_{x}} &= \frac{\lfloor\mu_{x} \cdot (2^{16}-1) + 2^{-1}\rfloor}{2^{16}-1} \\
  \hat{\sigma_{x}} &= \frac{\lfloor\sigma_{x} \cdot (2^{16}-1) + 2^{-1}\rfloor}{2^{16}-1}
\end{align}
Here, $\hat{x}^{c}$ represents the color-corrected reconstruction, and $\mu_{x}$ and $\sigma_{x}$ contain the mean and variance values for the RGB channels, each represented as 32-bit floating point values. We find that quantizing these values to 16 bits does not significantly affect correction performance. This strategy effectively refines the color of high-resolution reconstructions with only a minimal increase in bit cost (96 bits per image), as demonstrated in Fig. \ref{color}.

\section{Network Structure}

\begin{figure*}[!ht]
    \centering
    \includegraphics[width=\textwidth]{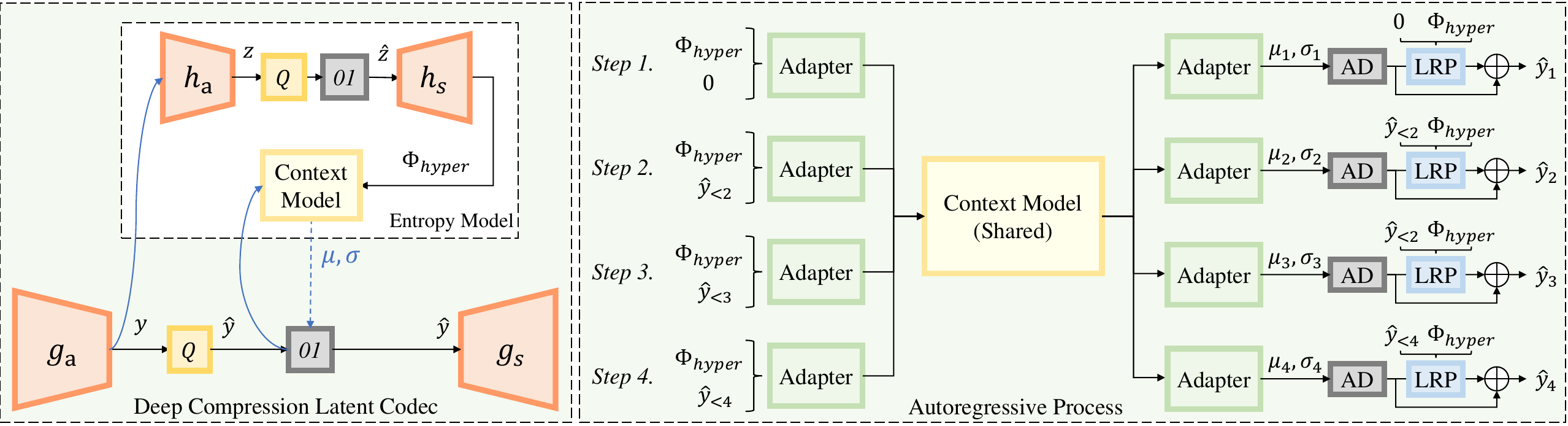}  
    \caption{\textbf{(Left) Illustration of the entropy model.} We build our entropy model on the basis of \cite{minnen2018joint}, which consists of a pair of hyper transforms, $h_a$ and $h_s$, and a context model to perform entropy estimation for $\hat{y}$ in an autoregressive manner. \textbf{(Right) Illustration of the 4-step autoregressive process.} We divide $\hat{y}$ into 4 groups ($\hat{y}_1$, $\hat{y}_2$, $\hat{y}_3$ and $\hat{y}_4$) using quadtree partition \cite{li2023neural}. For each $\hat{y}_i$, we estimate its Gaussian parameters, $\mu_i$ and $\sigma_i$, with the hyperprior $\Phi_{hyper}$ and previously decoded groups $\hat{y}_{<i}$. The parameter networks contain a shared context model and private adapters. AD represents arithmetic decoding the bitstream of $\hat{y}_i$ given corresponding Gaussian parameters, $\mu_i$ and $\sigma_i$. Additionally, we incorporate latent residual prediction (LRP) \cite{minnen2020channel} to alleviate the quantization error.}
    \label{network1}
\end{figure*}

\begin{figure*}[!ht]
    \centering
    \includegraphics[width=\textwidth]{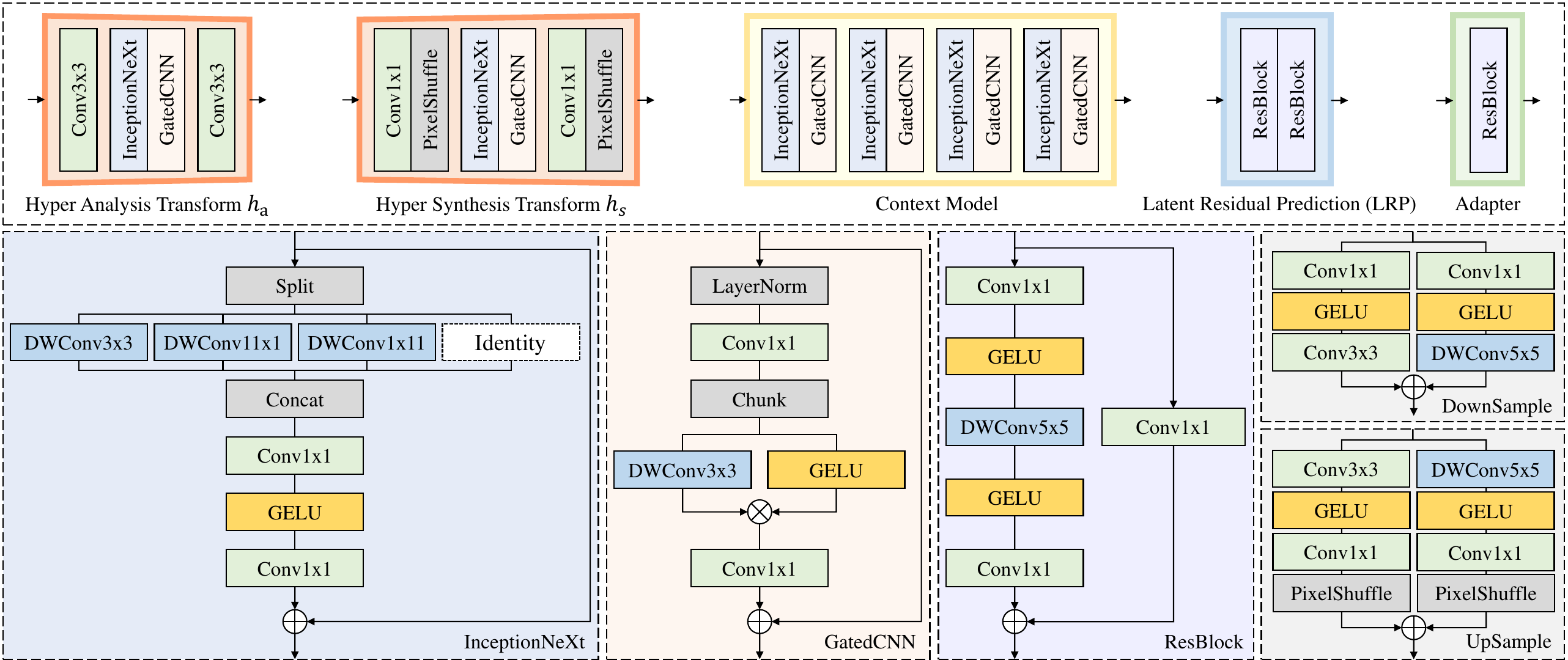}  
    \caption{\textbf{Module structures and network details.}}
    \label{network2}
\end{figure*}

We present our entropy model in Fig. \ref{network1}, with the detailed network architecture shown in Fig. \ref{network2}. Given the quantized latent $\hat{y}$, the entropy model estimates its distribution for arithmetic coding. Following \cite{minnen2018joint}, our entropy model is built with a hyperprior module and an autoregressive context model, where we first obtain and transmit a hyperprior $\Phi_{hyper}$ from $y$ using the hyper transform $h_a$ and $h_s$:
\begin{equation}
\begin{aligned}
  z=h_{a}(y), \hat{z}=Q(z), \Phi_{hyper}=h_{s}(\hat{z})
\end{aligned}
\end{equation}
Here, $y$ has 320 channels with $64\times$ (a spatial compression ratio of 64), while $z$ and $\hat{z}$ have 160 channels with $256\times$. To balance the coding performance and efficiency, we construct a 4-step autoregressive process using quadtree partition \cite{li2023neural} and latent residual prediction \cite{minnen2020channel}. The detailed autoregressive process to estimate the Gaussian parameters, $\mu$ and $\sigma$, for $\hat{y}$ is illustrated in Fig. \ref{network1}. Following this, arithmetic coding is applied to encode $\hat{y}$ into a bitstream, or decode $\hat{y}$ from the bitstream. For efficient network construction, we primarily rely on modified versions of InceptionNeXt \cite{yu2024inceptionnext} and GatedCNN \cite{yu2024mambaout}, as detailed in Fig. \ref{network2}.

\begin{table*}[!t]
    \centering
    \begin{tabular}{c|cccc|ccccc}
        \Xhline{1.0pt}
        \multirow{2}{*}{Method} & \multicolumn{4}{c|}{Encoding Time (s)}                                      & \multicolumn{5}{c}{Decoding Time (s)} \\
        \cline{2-10}
        ~      & $\mathcal{E}_{\mathrm{SD}}$ & $\mathcal{E}_{\mathrm{Aux}}$           & $g_a$        & EE  & ED & $g_s$         & $\mathcal{D}_{\mathrm{Aux}}$           & $\epsilon_{\mathrm{SD}}$           & $\mathcal{D}_{\mathrm{SD}}$ \\
        \hline
        \hline
        StableCodec (Ours) & 0.108 & 0.014 & 0.005 & 0.029 & 0.041 & 0.004 & 0.004 & 0.112 & 0.161 \\
        ELIC \cite{he2022elic} & - & - & 0.015 & 0.138 & 0.230 & 0.016 & - & - & - \\
        \Xhline{1.0pt}
    \end{tabular}
    \caption{\textbf{Runtime analysis of specific modules in seconds} averaged on Kodak \cite{kodak}. $\mathcal{E}_{\mathrm{SD}}$ and $\mathcal{D}_{\mathrm{SD}}$ represent the VAE encoder and decoder of SD-Turbo, while EE and ED denote entropy encoding and decoding with the entropy model. We add representative neural codec ELIC \cite{he2022elic} for comparison, which only contains the analysis transform $g_a$, the synthesis transform $g_s$ and the entropy model.}
    \label{runtime}
\end{table*}

\begin{figure*}[!ht]
    \centering
    \includegraphics[width=\textwidth]{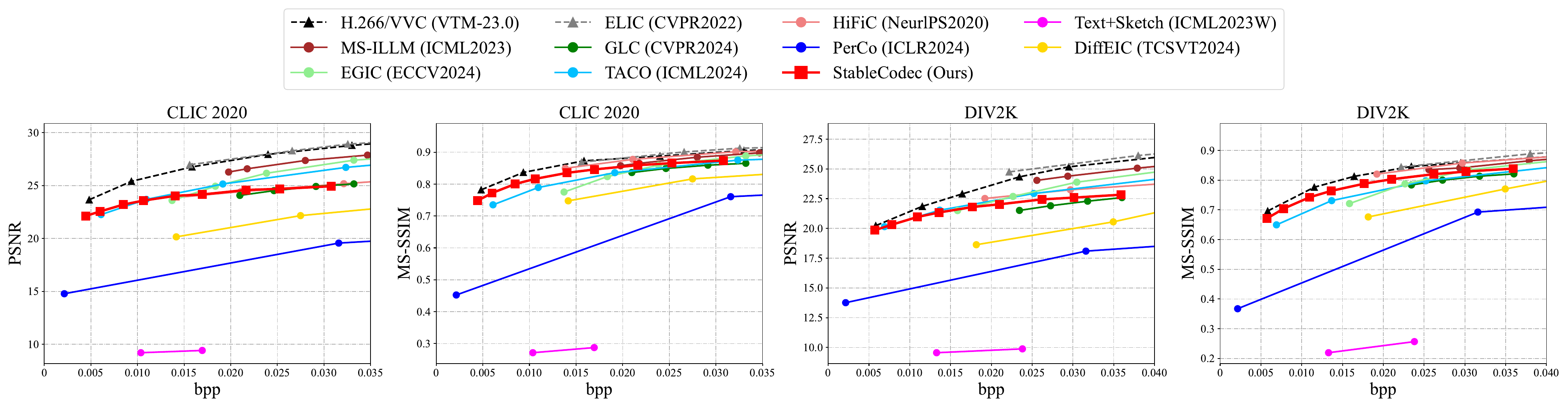}  
    \caption{\textbf{Additional rate-distortion curves on CLIC 2020 \cite{toderici2020clic} and DIV2K \cite{agustsson2017ntire} in terms of PSNR and MS-SSIM.}}
    \label{rdcurve}
\end{figure*}

\begin{table*}[!t]
    \centering
    \begin{tabular}{c|cccccccc}
        \Xhline{1.0pt}
        Method & HiFiC & MS-ILLM & Text+Sketch & PerCo & DiffEIC & EGIC & TACO & StableCodec (Ours) \\
        \hline
        \hline
        Bitrate (bpp) & 0.0268 & 0.0262 & 0.0274 & 0.0321 & 0.0375 & 0.0247 & 0.0258 & 0.0250 \\
        \hline
        Top-1 Votes & 20 & 26 & 11 & 24 & 43 & 29 & 54 & \textbf{513} \\
        Percentage & 2.78\% & 3.61\% & 1.53\% & 3.33\% & 5.97\% & 4.03\% & 7.50\% & \textbf{71.25\%} \\
        \Xhline{1.0pt}
    \end{tabular}
    \caption{\textbf{Top-1 user preference.} We evaluate reconstructions from different methods at similar ultra-low bitrates using the Kodak dataset \cite{kodak}. Our study involves 30 participants, yielding a total of 720 evaluated cases. In each case, we display the ground-truth image alongside eight reconstructions from different methods, and invite participates to select the most ``consistent" one compared with the ground-truth.}
    \label{us}
\end{table*}

\section{Runtime Analysis}

We conduct detailed runtime analysis of different modules in StableCodec using a single RTX 3090 GPU, and display the results in Table \ref{runtime}. Specifically, we examine the time consumption of the VAE encoder $\mathcal{E}_{\mathrm{SD}}$, auxiliary encoder $\mathcal{E}_{\mathrm{Aux}}$, $g_a$ and entropy encoding during the encoding process, and those of the entropy decoding, $g_s$, auxiliary decoder $\mathcal{D}_{\mathrm{Aux}}$, one-step denoising Unet $\epsilon_{\mathrm{SD}}$ and VAE decoder $\mathcal{D}_{\mathrm{SD}}$ during the decoding process. For comparison, we add the representative VAE-based neural codec ELIC \cite{he2022elic}, which only contains $g_a$, $g_s$ and the entropy model.

Since we use the analysis transform $g_a$ of a pre-trained ELIC model to serve as $\mathcal{E}_{\mathrm{Aux}}$, the time consumption of ``StableCodec - $\mathcal{E}_{\mathrm{Aux}}$" is close to that of ``ELIC - $g_a$". Besides, the time consumption of $g_a$, $g_s$ and entropy coding in StableCodec is much smaller than those of ELIC. This is because StableCodec adopts Deep Compression Latent Codec with advanced 4-step autoregressive entropy model and network designs, performing efficient transform coding at $16\times$ and entropy estimation at $64\times$, while ELIC performs transform coding on original images and entropy estimation at $16\times$. Benefit from these designs, StableCodec is able to achieve comparable coding speed with mainstream neural codecs, significantly outperforms existing diffusion-based methods as suggested in Table 2.

\section{User Study}

To provide a more comprehensive evaluation of reconstruction quality at ultra-low bitrates, we conduct a user study on the Kodak dataset \cite{kodak} using a top-1 user preference approach. We compare StableCodec against seven representative generative image codecs: HiFiC \cite{mentzer2020high}, MS-ILLM \cite{muckley2023improving}, Text+Sketch \cite{lei2023text+}, PerCo \cite{careil2023towards}, DiffEIC \cite{li2024towards}, EGIC \cite{korber2024egic}, and TACO \cite{lee2024neural}, all evaluated at similar average bitrates. To produce the reconstructions, we use the official weights of Text+Sketch, PerCo (SD) \cite{korber2024perco} and DiffEIC, while HiFiC, MS-ILLM, EGIC and TACO are either re-trained or finetuned from existing weights to reach specific bitrates.

Each participant in our study examines 24 cases, requiring an average of three minutes to complete. For each case, we present a ground-truth image alongside eight reconstructions from different methods, displayed in 2 rows and 4 columns with random order. Participants are asked to select the reconstruction they find most ``consistent" with the ground-truth image. A total of 30 participants completed the study, yielding 720 evaluated cases. The results, summarized in Table \ref{us}, show that StableCodec reconstructions were preferred in over 70\% of cases, demonstrating its superior visual consistency as perceived by human observers.

\section{Visual Performance}

\begin{figure*}[!t]
    \centering
    \includegraphics[width=\textwidth]{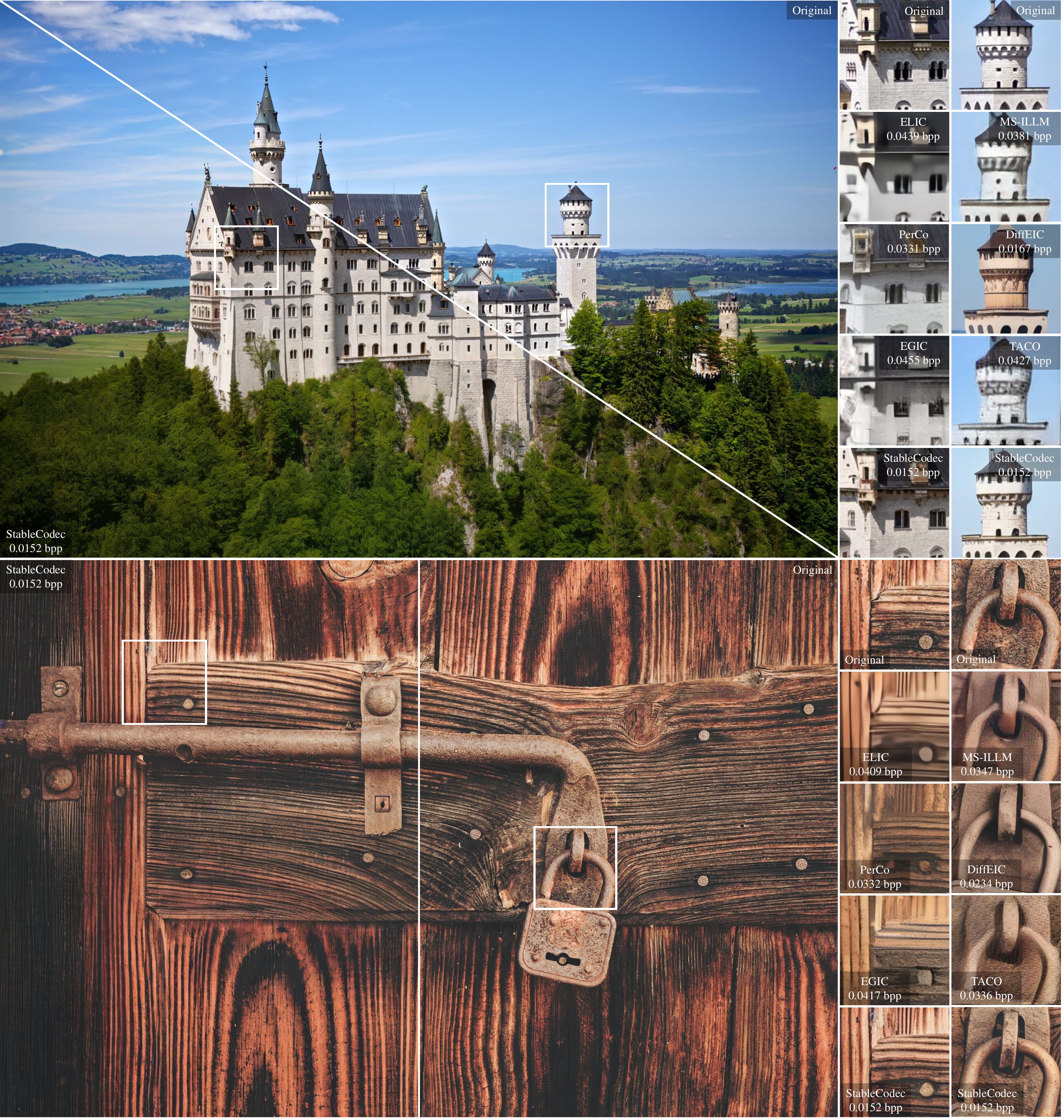}  
    \caption{\textbf{Visual examples and comparisons on 2K-resolution images from DIV2K.}}
    \label{2KDIV2K}
\end{figure*}

\begin{figure*}[!t]
    \centering
    \includegraphics[width=\textwidth]{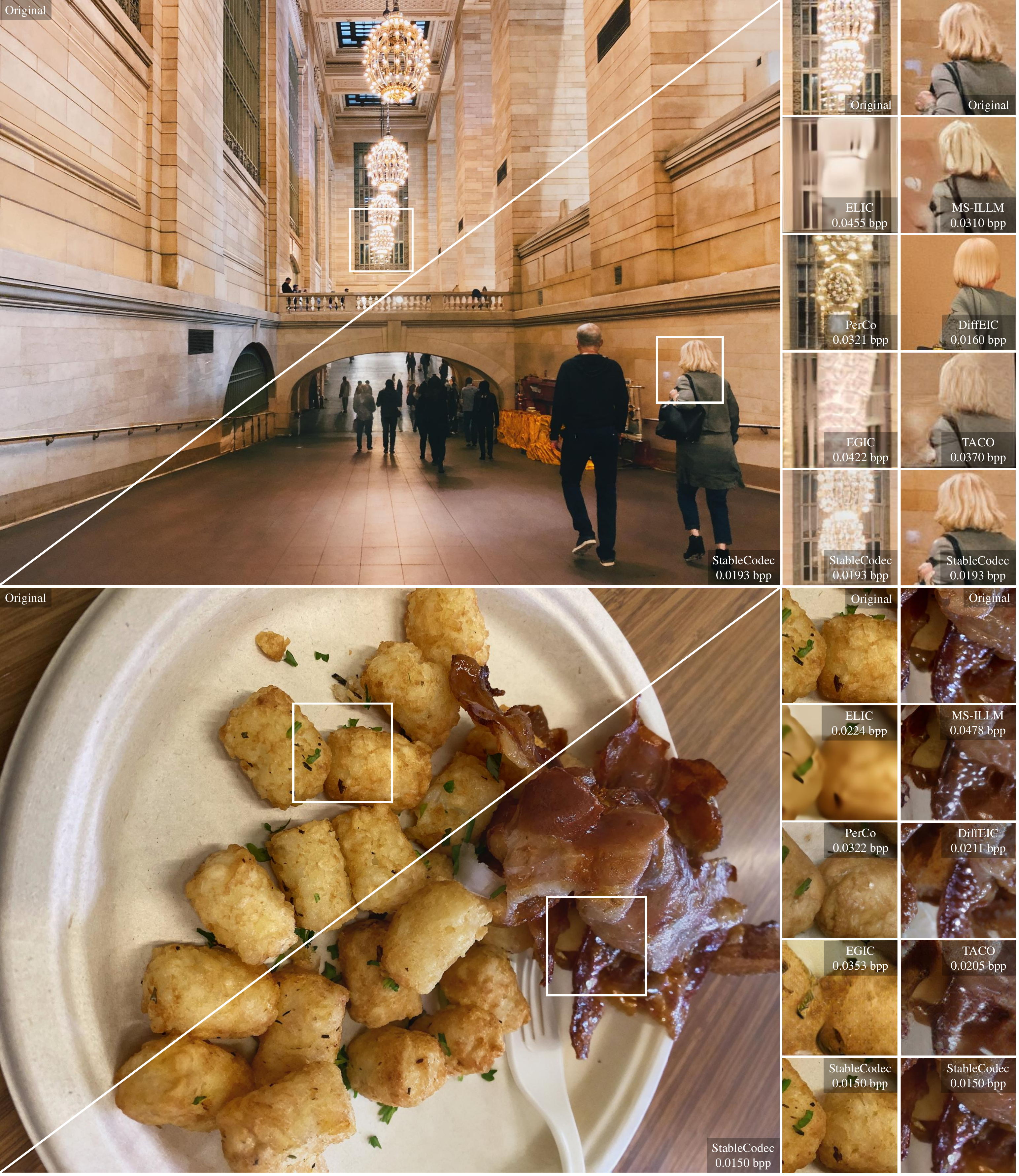}  
    \caption{\textbf{Visual examples and comparisons on 2K-resolution images from CLIC 2020.}}
    \label{2KCLIC}
\end{figure*}

\begin{figure*}[!t]
    \centering
    \includegraphics[width=\textwidth]{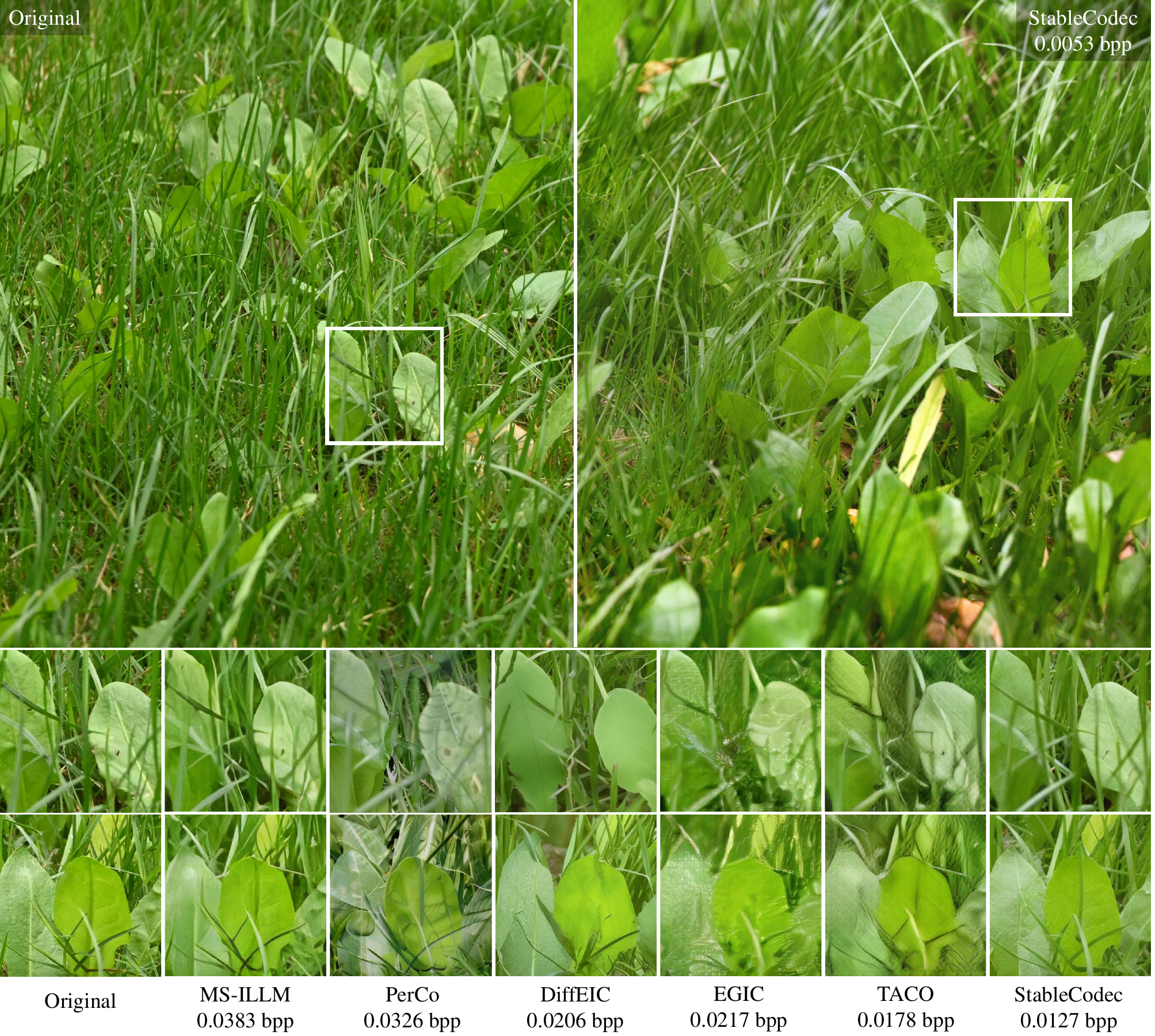}  
    \caption{\textbf{Visual examples and comparisons on 4K-resolution images from USTC-TD \cite{li2024ustc}.}}
    \label{4K1}
\end{figure*}

\begin{figure*}[!t]
    \centering
    \includegraphics[width=\textwidth]{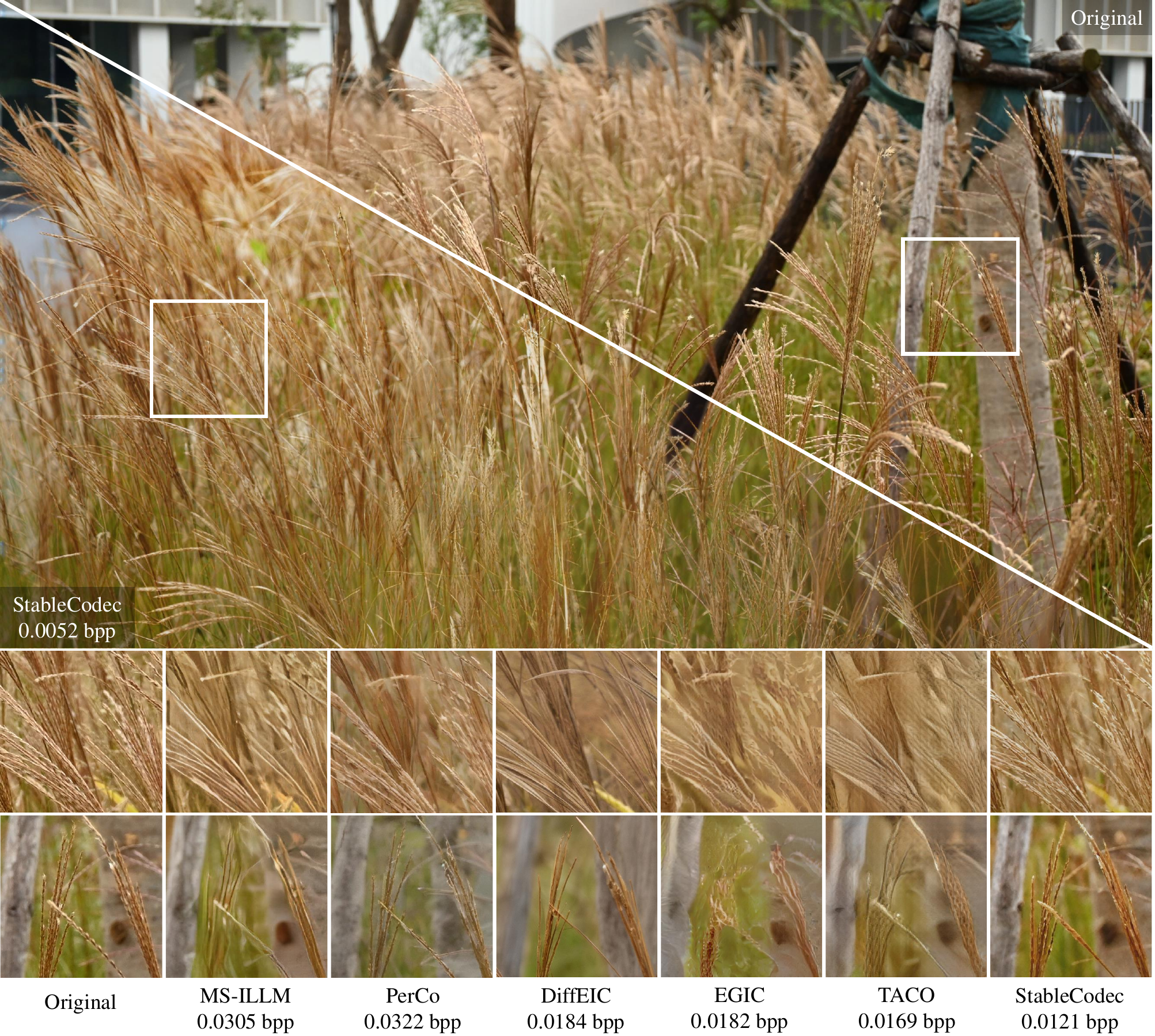}  
    \caption{\textbf{Visual examples and comparisons on 4K-resolution images from USTC-TD \cite{li2024ustc}.}}
    \label{4K2}
\end{figure*}

In this section, we display more visual examples and comparisons on high-quality images from DIV2K \cite{agustsson2017ntire} (Fig. \ref{2KDIV2K}), CLIC 2020 \cite{toderici2020clic} (Fig. \ref{2KCLIC}) and USTC-TD \cite{li2024ustc} (Fig. \ref{4K1} and Fig. \ref{4K2}). We compare the proposed StableCodec with existing methods, including ELIC \cite{he2022elic}, MS-ILLM \cite{muckley2023improving}, PerCo \cite{careil2023towards}, EGIC \cite{korber2024egic}, DiffEIC \cite{li2024towards}, and TACO \cite{lee2024neural}, all at ultra-low bitrates. Notably, StableCodec outperforms the competing methods in terms of both semantic consistency and textual realism, while consuming fewer bits.

\section{Quantitative Results}

In Fig. \ref{rdcurve}, we provide additional PSNR and MS-SSIM comparisons on CLIC 2020 and DIV2K as a supplement for Fig. 6. As discussed in Section 4.1, pixel-level metrics like PSNR, MS-SSIM, and LPIPS have notable limitations \cite{jia2024generative, lei2023text+, ding2020image, careil2023towards} due to their emphasis on pixel accuracy rather than semantic consistency or textual realism, making them less suitable for evaluating ultra-low bitrate compression. Therefore, for StableCodec, we primarily focus on FID, KID, and DISTS, which offer a more accurate assessment of quality in severely compressed images.

\end{document}